\documentclass[prd,preprintnumbers,superscriptaddress,nofootinbib,showpacs,onecolumn]{revtex4-2}
\usepackage[toc,page]{appendix}
\usepackage{pstricks}
\usepackage{color}
\usepackage{amssymb,amsmath,bm,bbm}
\usepackage{epsf}
\usepackage{epsfig}
\usepackage{afterpage}
\usepackage{longtable}
\usepackage{latexsym, mathrsfs}
\usepackage{graphics}
\usepackage{url}
\usepackage{paralist}
\usepackage{bbold}

\newcommand{\be}{\begin{equation}}
\newcommand{\ee}{\end{equation}}
\newcommand{\bea}{\begin{eqnarray}}
\newcommand{\eea}{\end{eqnarray}}
\newcommand{\nn}{\nonumber}
\newcommand{\dd}{\displaystyle}

\newcommand{\epsK}{\varepsilon_K}
\newcommand{\gev}{\, \rm GeV}
\newcommand{\mev}{\, \rm MeV}
\begin{document}

\preprint{BARI-TH/21-728}

\title{ $c \to u \nu {\bar \nu}$  transitions of $B_c$ mesons: \\  331 model facing  Standard Model null tests }
\author{Pietro~Colangelo}
\email[Electronic address:]{pietro.colangelo@ba.infn.it} 
\affiliation{Istituto Nazionale di Fisica Nucleare, Sezione di Bari, via Orabona 4, 70126 Bari, Italy}
\author{Fulvia~De~Fazio}
\email[Electronic address:]{fulvia.defazio@ba.infn.it} 
\affiliation{Istituto Nazionale di Fisica Nucleare, Sezione di Bari, via Orabona 4, 70126 Bari, Italy}
\author{Francesco~Loparco}
\email[Electronic address:]{francesco.loparco1@ba.infn.it} 
\affiliation{Istituto Nazionale di Fisica Nucleare, Sezione di Bari, via Orabona 4, 70126 Bari, Italy}
\affiliation{Dipartimento Interateneo di Fisica "Michelangelo Merlin", Universit\`a degli Studi di Bari, via Orabona 4, 70126 Bari, Italy}

\begin{abstract}
The Glashow-Iliopoulos-Maiani mechanism is extremely efficient to suppress the
flavor-changing neutral current decays of charmed hadrons induced by the $c \to u$ transitions, making such processes particularly  sensitive to  phenomena beyond the Standard Model. 
In particular, $c \to u$ decays with a neutrino pair in the final state are theoretically appealing due to the small long-distance contributions. Moreover, in the framework of the Standard Model effective field theory (SMEFT),  the  $SU(2)_L$ invariance allows us to relate the Wilson coefficients in the effective Hamiltonian governing  the $c \to u \nu {\bar \nu}$ decays to the coefficients in the $c \to u \ell^+ \ell^-$ Hamiltonian.
We analyze the $B_c \to B^{(*)+} \nu {\bar \nu}$ decays, for which  branching fractions of at most  ${\cal O}(10^{-16})$ are predicted in the Standard Model including short- and long-distance contributions, so small that they can be considered as  null tests.   Using SMEFT and the
 relation to the $c \to u \ell^+ \ell^-$  processes we study the largest enhancement achievable in generic new physics scenarios, then we focus  on a particular extension of the Standard Model, the 331 model.  SMEFT relations and the connection with  $c \to u \ell^+ \ell^-$ imply that ${\cal B}(B_c \to B^{(*)+} \nu {\bar \nu})$ could  even reach ${\cal O}(10^{-6})$, an extremely large enhancement.  A  less pronounced  effect is found in the 331 model, with  ${\cal O}(10^{-11})$ predicted branching fractions. Within the 331 model  correlations  exist among  the $B_c \to B^{(*)+}\nu \bar \nu$ and  $K\to \pi \nu \bar \nu$,  $B\to (X_s, K, K^*) \nu \bar \nu$  channels.
\end{abstract}

\maketitle

\section{Introduction}
In the Standard Model (SM) the flavor-changing neutral current (FCNC) transitions occur  at loop level and are generally characterized by Cabibbo-Kobayashi-Maskawa (CKM)  and loop suppressions. The CKM cancellation mechanism is particularly efficient in the processes involving up-type quarks  which take place through penguin and box diagrams with internal down-type quark exchanges.  This is the case of the charmed hadron decays induced by  the $c \to u \ell^+ \ell^-$ and $c \to u \nu {\bar \nu}$ transitions, for which tiny branching fractions are predicted in SM  considering the short-distance amplitude \cite{Gisbert:2020vjx}. The modes with  charged dileptons are polluted by long-distance (LD) hadronic contributions,  and the phase-space regions where such terms are large must be cut  to pin down the effects of the short-distance term \cite{Burdman:2001tf}.
In the $c \to u$ dineutrino modes long-distance effects are smaller than in the charged dilepton modes.  Therefore, such processes  represent genuine null tests of the SM: their observation would be an indication of phenomena beyond the Standard Model (BSM). Among all hadrons, the decays of $B_c$  induced by the  $c \to u$ transitions are particularly interesting, since in this case the main long-distance contributions affect a region of the phase-space near the end point, differently, e.g., from $D$, $D_s$ and $\Lambda_c$. Hence, $B_c$  plays an  important  role in testing the Standard Model \cite{Colangelo:2021dnv}. On general grounds, searching for new physics (NP) effects requires the analysis of several modes induced by the same underlying transition: the correlations among the various observables are important to identify the possible NP contributions and to relate them to the structure of the SM extensions.

The  short-distance low-energy Hamiltonian  governing the $c \to u \nu {\bar \nu}$  transition has a simple structure.  For left-handed neutrinos it consists of  two operators,
\be
{\cal H}_{eff} = C_L \; {\cal Q}_L + \, C_R \; {\cal Q}_R \, ,
\label{hamil}
\ee
with 
\bea
{\cal Q}_{L} &=& ({\bar u} \gamma^\mu ( 1 - \gamma_5) c) ({\bar \nu} \gamma_\mu ( 1- \gamma_5) \nu ) \nn \\
{\cal Q}_{R} &=& ({\bar u} \gamma^\mu ( 1 + \gamma_5) c) ({\bar \nu} \gamma_\mu ( 1- \gamma_5) \nu ) . \label{eq:OLR}
\eea
 In  the SM the Hamiltonian comprises only ${\cal Q}_L$. 
 The Wilson coefficient $C_L^{SM}$  is obtained from  loop diagrams with down-type quark exchanges,
 \be
 C_L^{SM}=-\displaystyle\frac{G_F}{\sqrt{2}}\frac{\alpha}{2\pi \sin^2 \theta_W}\sum_{q=d,s,b} \lambda^{q} X(x_q) . \label{eq:CLSM}
 \ee
  In \eqref{eq:CLSM}
 $G_F$ is the Fermi constant, $\alpha$ the fine structure constant, $\theta_W$  the Weinberg angle,  $\lambda^{q}=V_{cq}^* \,V_{uq}$ with $V_{ij}$ the CKM matrix elements. The Inami-Lim function $X(x_q)$,  depending on  $x_q=m_q^2/M_W^2$, can be found in \cite{Buras:2020xsm}. The dominant contribution from the  intermediate $b$ quark  provides $|C_L^{SM}|\simeq {\cal O}(10^{-13})$.
 Analogously, the transitions  $s \to d \nu {\bar \nu}$ and $b \to (s,d) \nu {\bar \nu}$
are governed by a low-energy Hamiltonian with the  structure of \eqref{hamil} and intermediate up-type quarks. 
BSM phenomena can manifest  themselves through the enhancement of $C_L$ and through the effects of the operator ${\cal Q}_R$.

FCNC dineutrino modes have been extensively  studied in the case of  strange and beauty quarks.  The 
 $K^+ \to \pi^+ \nu {\bar \nu}$ and $K_L \to \pi^0 \nu {\bar \nu}$ transitions are under strict theoretical control \cite{Buras:2020xsm} and intense experimental scrutiny \cite{NA62:2021zjw,Ahn:2018mvc,Ahn:2020opg}. In the beauty sector, the modes $B \to K^{(*)} \nu \bar \nu$ have been theoretically investigated \cite{Colangelo:1996ay,Buchalla:2000sk,Altmannshofer:2009ma,Biancofiore:2014uba,Buras:2014fpa,Calibbi:2015kma,Das:2017ebx,Ahmady:2018fvo,Descotes-Genon:2020buf,Browder:2021hbl,He:2021yoz,Bause:2021ply}  and are within the reach of the present facilities \cite{BaBar:2013npw,Blake:2016olu,Grygier:2017tzo,Belle-II:2021rof}.

As for the charm sector, a few  studies  have analyzed the FCNC dineutrino modes in the  SM and BSM frameworks \cite{Bause:2020auq,Bause:2020xzj,Faisel:2020php,Fajfer:2021woc}.
Here, we focus on $B_c \to B^{(*)+} \nu {\bar \nu}$ decays, for which  lattice QCD results for  the hadronic  form factors can be used \cite{Cooper:2020wnj},  with a  control of the theoretical uncertainty related to nonperturbative QCD quantities.
From the experimental point of view, these modes will be accessible at high energy $e^+e^-$ colliders, namely  the  planned future circular collider FCC-ee  machine running at  the $Z^0$ peak.

We proceed both in a model-independent way and in a defined BSM framework. In the next section we apply the Standard Model effective field theory (SMEFT) to  relate the Wilson coefficients in the low-energy $c \to u \nu {\bar \nu}$  Hamiltonian \eqref{hamil}  to the coefficients  in the $c \to u \ell^+ \ell^-$ Hamiltonian, as done in  \cite{Bause:2020auq,Bause:2020xzj}.   This allows us to establish  the largest  enhancement for the $B_c$ branching fractions achievable in a generic NP scenario, with the numerical  results  discussed in Sec. \ref{sec:modes}. In Sec. \ref{sec:331} we focus on a definite NP model, the  331 model  in four  variants. We observe that  in this framework it is possible to relate  the charm to  the strange and beauty  quark sectors, and that the $c \to u$ processes can be constrained  using bounds from  $\Delta S=2$ and $\Delta B=2$  observables. The correlations among     $B_c$ and kaon and $B$ meson dineutrino  decays are described in   Sec. \ref{sec:corr}. In the last section we draw our conclusions.  

\section{Relating the $c \to u$ dineutrino and charged dilepton   modes using  SMEFT}\label{sec:two}
A relation between the $c \to u$ dineutrino and the  $c \to u$ charged dilepton modes can be established on the basis of $SU(2)_L$ invariance
using the Standard Model effective theory   \cite{Bause:2020auq,Bause:2020xzj}.  
 Considering the possibility of lepton flavor violation, one focuses on  
 $c \to u \nu_i {\bar \nu}_j$  transitions, with the indices $i,j$   denoting the neutrino flavors. 
 The coefficients $C_{L}$ and $C_R$  in the low-energy Hamiltonian \eqref{hamil} become
lepton-flavor dependent $C_{L,R}^{i,j}$ and  can be combined giving
\be
x_U^\pm=\sum_{i,j=1,2,3} |{\tilde C}_L^{i,j} \pm {\tilde C}_R^{i,j}|^2 \,\,  \label{xUpiu} 
\ee
and
\be
x_U=\frac{x_U^+ + x_U^-}{2}\,\, , \label{xU}
\ee
  with ${\tilde C}_{L,R}$  defined by
  $C_{L,R}=-\displaystyle\frac{G_F}{\sqrt{2}}\frac{\alpha}{4\pi}{\tilde C}_{L,R}$.
The combinations \eqref{xUpiu} and \eqref{xU} account for  the contributions of both the operators  ${\cal Q}_L$ and ${\cal Q}_R$.

The relation of $C_{L,R}$ to the Wilson coefficients in the $c \to u \ell^+ \ell^-$  low-energy Hamiltonian has been proposed in \cite{Bause:2020auq,Bause:2020xzj}. For two generic quarks $q_1\,$  and $\,q_2$ the $q_1 \to q_2 \ell^+ \ell^-$ general Hamiltonian reads \cite{Aebischer:2015fzz}:

\be
H_{eff}^{q_1 \to q_2 \ell^+ \ell^-}=-4 \frac{G_F}{\sqrt{2}}\left[\sum_{i=9,10,S,P} \Big(C_i {\cal Q}_i +C_i^\prime {\cal Q}_i^\prime \Big) +C_T {\cal Q}_T+C_{T5} {\cal Q}_{T5} \right], \label{Heffmumu}
\ee
with the operators
\bea
{\cal Q}_{9}&=& \frac{\alpha}{4 \pi}  ({\bar q_2} \gamma_\mu P_L \, q_1)({\bar \ell} \gamma^\mu  \ell) \quad \quad \qquad \hspace{0.15cm}
{\cal Q}_{9}^\prime=  \frac{\alpha}{4 \pi}  ({\bar q_2} \gamma_\mu P_R \, q_1)({\bar \ell} \gamma^\mu  \ell) \nn \\
{\cal Q}_{10}&=& \frac{\alpha}{4 \pi}  ({\bar q_2} \gamma_\mu P_L \, q_1)({\bar \ell} \gamma^\mu \gamma_5 \ell) \quad \qquad
{\cal Q}_{10}^\prime = \frac{\alpha}{4 \pi}  ({\bar q_2} \gamma_\mu P_R \,q_1)({\bar \ell} \gamma^\mu \gamma_5 \ell) \nn  \\
{\cal Q}_S&=&   ({\bar q_2} P_R \,q_1)({\bar \ell}   \ell) \qquad \qquad \qquad \quad \hspace{0.2cm}
{\cal Q}_S^\prime=   ({\bar q_2} P_L \,q_1)({\bar \ell}   \ell)  \nn \\
{\cal Q}_P&=&   ({\bar q_2} P_R \,q_1)({\bar \ell}  \gamma_5 \ell)  \qquad \qquad \qquad \hspace{0.25cm}
{\cal Q}_P^\prime=  ({\bar q_2} P_L \,q_1)({\bar \ell}  \gamma_5 \ell)   \\
{\cal Q}_T&=&
({\bar q_2} \sigma_{\mu\nu}  \, q_1)({\bar \ell} \sigma^{\mu \nu}\ell) 
\quad \nn \\
{\cal Q}_{T5}&=&
 ({\bar q_2} \sigma_{\mu\nu} \, q_1)({\bar \ell} \sigma^{\mu \nu} \gamma_5 \ell) \,\,   \nn 
\eea
and $\dd P_{R,L}=\frac{1\pm\gamma_5}{2}$.
The relations are obtained using   the SMEFT  operators  classified in Ref.~\cite{Grzadkowski:2010es}. The   tree-level matching of  the  dimension-6 four-fermion  operators invariant under the SM  $SU(3)_C \times SU(2)_L \times U(1)_Y$ gauge group with the  Hamiltonian   (\ref{hamil})  gives the relations:
\bea
 C_L^{c \to u}&=&\frac{v^2}{2 \Lambda^2} \Big[\left(C_{lq}^{(1)}+C_{lq}^{(3)} \right) +\left(C_{\varphi q}^{(1)}-C_{\varphi q}^{(3)} \right) \Big]\nn \\
 C_R^{c \to u}&=&\frac{v^2}{2 \Lambda^2} \Big( C_{lu}+C_{\varphi u } \Big) . \label{smeftneutrini}
\eea
$C_{L,R}^{c \to u}$ are defined by
$C_{L,R}=-\displaystyle\frac{G_F}{\sqrt{2}} C_{L,R}^{c \to u}$, with  $C_{L,R}$   in ~(\ref{hamil}).
The  operators corresponding to the coefficients in the rhs of Eq.~\eqref{smeftneutrini}  are expressed in the Warsaw basis  \cite{Grzadkowski:2010es}.
   In this equation $v$ is the electroweak vacuum expectation value and $\Lambda$ the  matching scale of  NP with  the SMEFT.  
 
The relations between  the coefficients in the $c \to u \ell^+ \ell^-$  Hamiltonian \eqref{Heffmumu}  and  the coefficients of the SMEFT operators can also be worked out:
\bea
C_{9}^{c \to u}&=& \frac{\pi v^2}{\alpha \Lambda^2} \Big( C_{lq}^{(1)}-C_{lq}^{(3)}+C_{qe} \Big)+\frac{\pi v^2}{\alpha \Lambda^2}(-1+4 s_W^2)\left(C_{\varphi q}^{(1)}-C_{\varphi q}^{(3)} \right) 
\nn \\
C_{9}^{\prime \, c \to u}&=&\frac{\pi v^2}{\alpha \Lambda^2}\Big(  C_{eu}+C_{lu}+(-1+4 s_W^2)C_{\varphi u } \Big)
\nn \\
C_{10}^{c \to u}&=& -\frac{\pi v^2}{\alpha \Lambda^2} \Big( C_{lq}^{(1)}-C_{lq}^{(3)}-C_{qe} \Big)+\frac{\pi v^2}{\alpha \Lambda^2}\left(C_{\varphi q}^{(1)}-C_{\varphi q}^{(3)} \right) 
\nn \\
C_{10}^{\prime \, c \to u}&=& \frac{\pi v^2}{\alpha \Lambda^2}\Big(  C_{eu}-C_{lu}+C_{\varphi u } \Big)
\nn \\
C_S^{c \to u}&=&C_P^{c \to u}= -\frac{ v^2}{4 \Lambda^2}C_{lequ}^{(1)}
\label{smeftcarichi} \\
C_S^{\prime c \to u}&=&-C_P^{\prime c \to u}= -\frac{ v^2}{4 \Lambda^2}C_{lequ}^{(1)*}  \nn  \\
C_T^{c \to u}&=&-\frac{ v^2}{4 \Lambda^2}2{\rm Re}[C_{lequ}^{(3)}]
\nn  \\
C_{T5}^{c \to u}&=&-\frac{ v^2}{4 \Lambda^2}2i \,{\rm Im}[C_{lequ}^{(3)}] \,\,. \nn
\eea
The SMEFT operators have generation indices. The coefficients  in the rhs of  Eqs.~\eqref{smeftneutrini} and \eqref{smeftcarichi}  read $C=C_{ij12}$, with  $i$ and $j$  the lepton generation indices and
 $1,2$ indicating the u and c quark in  the first and  second generation.\footnote{  In Ref.\cite{Fuentes-Martin:2020lea} the relations in Eq.~\eqref{smeftcarichi} are  obtained  neglecting the contribution of the anomalous gauge boson couplings, which correspond to the SMEFT coefficients  with indices $\varphi$. 
The  relations for  the $c \to u$ transitions  are different from those  for $b \to s$   \cite{Aebischer:2015fzz}.}

The coefficients of the SMEFT operators appearing in the rhs of Eq.~(\ref{smeftneutrini})  are also comprised  in the rhs of Eq.~(\ref{smeftcarichi}).  This allows us to translate the experimental bounds  on the   $c \to u \ell^+ \ell^-$ modes, together with data on the  $s \to d \ell^+ \ell^-$ modes,  into an upper bound for the combination $x_U$ in Eq.~(\ref{xU}) \cite{Bause:2020auq,Bause:2020xzj}.
Indeed, the $SU(2)_L$ symmetry links the  $c \to u \ell^+ \ell^-$ with $c \to u \nu \bar \nu $ modes, and  the   $s \to d \ell^+ \ell^-$ with $c \to u \nu \bar \nu $ modes. The bound on $x_U$ is obtained  assuming conservatively  that the experimental limits on the charged dilepton  branching fractions  are saturated by the short-distance Hamiltonian \eqref{Heffmumu}. The limit  depends on additional assumptions  on the structure of the transitions;
the most stringent  one is  obtained assuming lepton universality (LU)  and  charged lepton flavor conservation (cLFC) \cite{Bause:2020auq,Bause:2020xzj}, with the results 
\bea
 x_U  &\le& x_U^{max} = 34  \hskip 1.1cm ({\rm LU}) \label{xULU}
 \\
x_U &\le& x_U^{max} = 196   \hskip 1cm ({\rm cLFC}) \label{xUcLFC} \,\, .
\eea

The  bounds  \eqref{xULU} and \eqref{xUcLFC}  have been considered in  the analysis of $D^{\pm}$,  $D^{0}$, $D_{s}$ and charmed baryon  decays induced by  $c \to u \nu \bar \nu$  \cite{Bause:2020auq,Bause:2020xzj}. Here we focus on  $B_c^+ \to B^{(*)+} \nu \bar \nu$. We use   the   lattice QCD  $B_c \to B_{d}$ form factors in \cite{Cooper:2020wnj}, and the $B_c \to B_{d}^*$ form factors derived in  \cite{Colangelo:2021dnv} applying 
 the heavy quark spin symmetry \cite{Jenkins:1992nb,Colangelo:1999zn}.
The $B_c^+ \to B^{(*)+}$ form factors  are obtained invoking the isospin symmetry.

\section{  $B_c^+ \to B^{(*)+} \nu {\bar \nu}$ decays}\label{sec:modes}
In the processes   $B_c^+(p) \to B^+(p^\prime) \nu(k_1) {\bar \nu}(k_2)$ and $B_c^+(p) \to B^{*+}(p^\prime ,\,\epsilon) \nu(k_1) {\bar \nu}(k_2)$ the  particle momenta   are $p,\,p^\prime, \, k_1,\,k_2$ and  $\epsilon$ is  the $B^*$ polarization vector.  Denoting by
 $E_{miss}$   the energy of the neutrino pair  in the $B_c$ rest frame, the dimensionless variable $x=\displaystyle\frac{E_{miss}}{m_{B_c}}$ varies in the  range   $\displaystyle{\frac{1-r}{2}} \le x \le 1-\sqrt{r}$,
with $\dd r=\frac{m_{B^{(*)+}}^2}{m_{B_c}^2}$.
The hadronic  matrix elements  in the decay amplitudes are parametrized in terms of form factors:
\bea
&&\langle B^+(p^\prime)| {\bar u} \gamma_\mu c| {B_c}(p) \rangle =   f_+(q^2) \Big(p_\mu+p_\mu^\prime  - \frac{m_{B_c}^2-m_{B}^2}{q^2} q_\mu\Big)+ \,f_0(q^2)\frac{m_{B_c}^2-m_{B}^2}{q^2} q_\mu  \quad \quad \label{BcP} 
\eea
and
\bea
&&\langle B^{*+}(p^\prime,\epsilon)|{\bar u} \gamma_\mu c| {B_c}(p) \rangle = 
- {2 V (q^2) \over m_{B_c}+m_{B^{*}}} i \epsilon_{\mu \nu \alpha \beta} \epsilon^{*\nu}  p^\alpha p^{\prime \beta}, \nn 
\\
&&\langle B^{*+}(p^\prime,\epsilon)|{\bar u} \gamma_\mu\gamma_5 c| {B_c}(p) \rangle =   (m_{B_c}+m_{B^{*}}) \Big( \epsilon^*_\mu -{(\epsilon^* \cdot q) \over q^2} q_\mu \Big) A_1(q^2) \label{BctoPstar}  \\
&&- {(\epsilon^* \cdot q) \over  m_{B_c}+m_{B^{*}}} \Big( (p+p^\prime)_\mu -{m_{B_c}^2-m_{B^{*}}^2 \over q^2} q_\mu \Big) A_2(q^2)  + (\epsilon^* \cdot q){2 m_{B^{*}} \over q^2} q_\mu A_0(q^2) \,\,.\nn \eea
The   $B_c^+ \to B^+ \nu {\bar \nu}$   missing energy distribution obtained from \eqref{hamil} involves  the   form factor $f_+(q^2)$:
\be
\frac{d \Gamma(B_c^+ \to B^+ \nu {\bar \nu})}{ dx} = 
3\;{ |C_L + C_R|^2 \,|f_+(q^2)|^2 \over 48 \pi^3 m_{B_c}} 
\lambda^{3/2}(q^2, m_{B_c}^2, m_{B}^2)\;,
\label{dGBu}
\ee
with $q=p-p^\prime$ and $\lambda$  the 
K\"all\'en function.
For   $B_c^+ \to B^{*+} \nu {\bar \nu}$ the missing energy  distributions for longitudinally and transversely polarized $B^*$ read
\bea
\frac{d \Gamma_L}{dx} &=& 3\,{ |C_L - C_R|^2 \over 24 \pi^3}
\frac{ |{\vec p}^\prime|}{ m_{B^{*}}^2} 
\Big( (m_{B_c} + m_{B^{*}})(m_{B_c} E^\prime-m_{B^{*}}^2) A_1(q^2) - \frac{2 m_{B_c}^2}{m_{B_c} + 
m_{B^{*}}} |{\vec p}^\prime|^2 A_2(q^2) \Big)^2 \nn \\
\frac{d \Gamma_\pm}{dx}&=&3 \frac{ |{\vec p}^\prime| q^2 }{ 24 \pi^3}
\left | (C_L + C_R) { 2 m_{B_c}|{\vec p}^\prime| \over m_{B_c} + m_{B^{*}}} V(q^2) \mp
(C_L - C_R) (m_{B_c} + m_{B^{*}}) A_1(q^2) \right |^2 ,
\label{BustarT}
\eea
with ${\vec p}^\prime$ and $E^\prime$  the $B^*$ three-momentum and energy in 
the $B_c$  rest frame. In Eqs.~(\ref{dGBu}) and (\ref{BustarT}) the relation
$q^2=m_{B_c}^2 (2 x- 1)+m_{B^{(*)}}^2$ is used;  the factor $3$ is due  to the sum over the  three neutrino flavors.

\begin{figure}[t]
\begin{center}
\includegraphics[width = 0.5\textwidth]{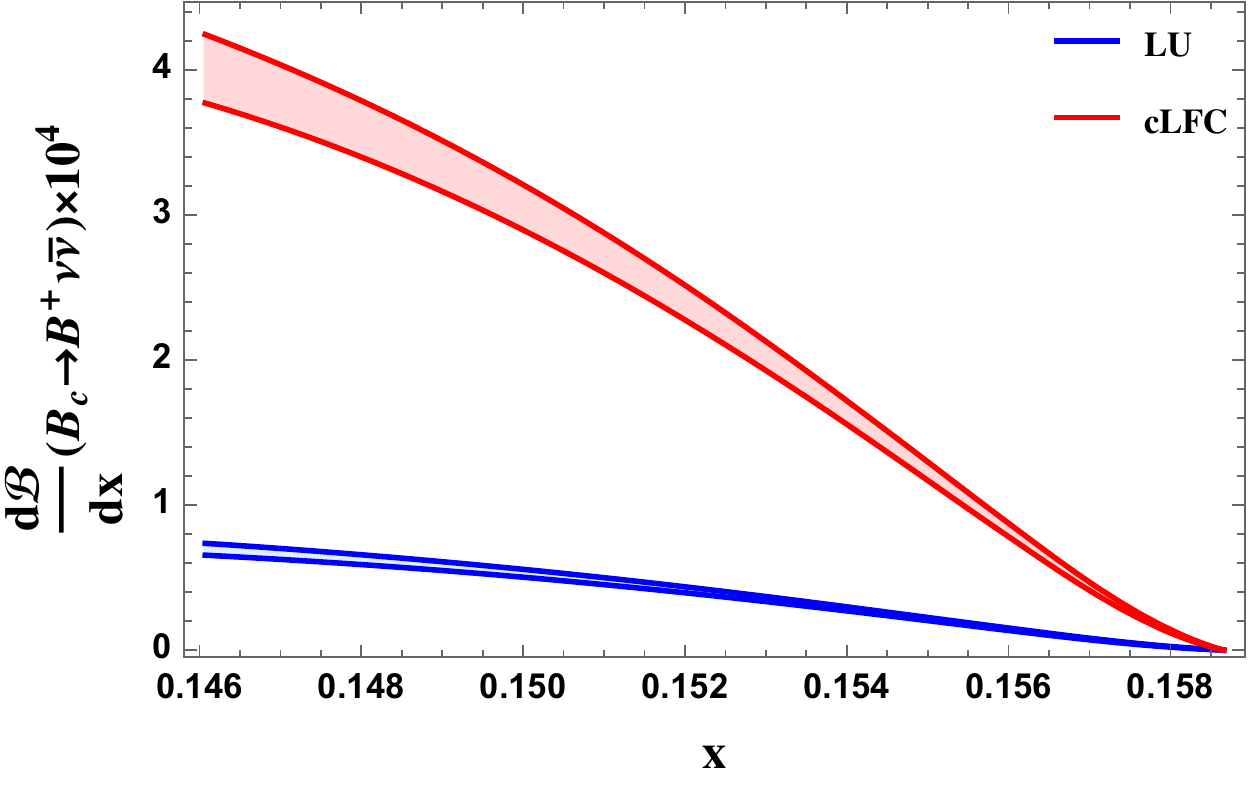}
    \caption{\small Missing energy distribution $d{\cal B}(B_c^+ \to B^+ \nu {\bar \nu})/dx$  for the largest value of the coefficients combination $x_U$ in Eqs.~(\ref{xULU}) (LU bound - blue curve) and (\ref{xUcLFC}) (cFLC bound - red curve). The widths of the curves  are obtained varying   the form factor parameters  \cite{Cooper:2020wnj,Colangelo:2021dnv}.
}\label{dBdxBu}
\end{center}
\end{figure}

\begin{figure}[b!]
\begin{center}
\includegraphics[width = 0.5\textwidth]{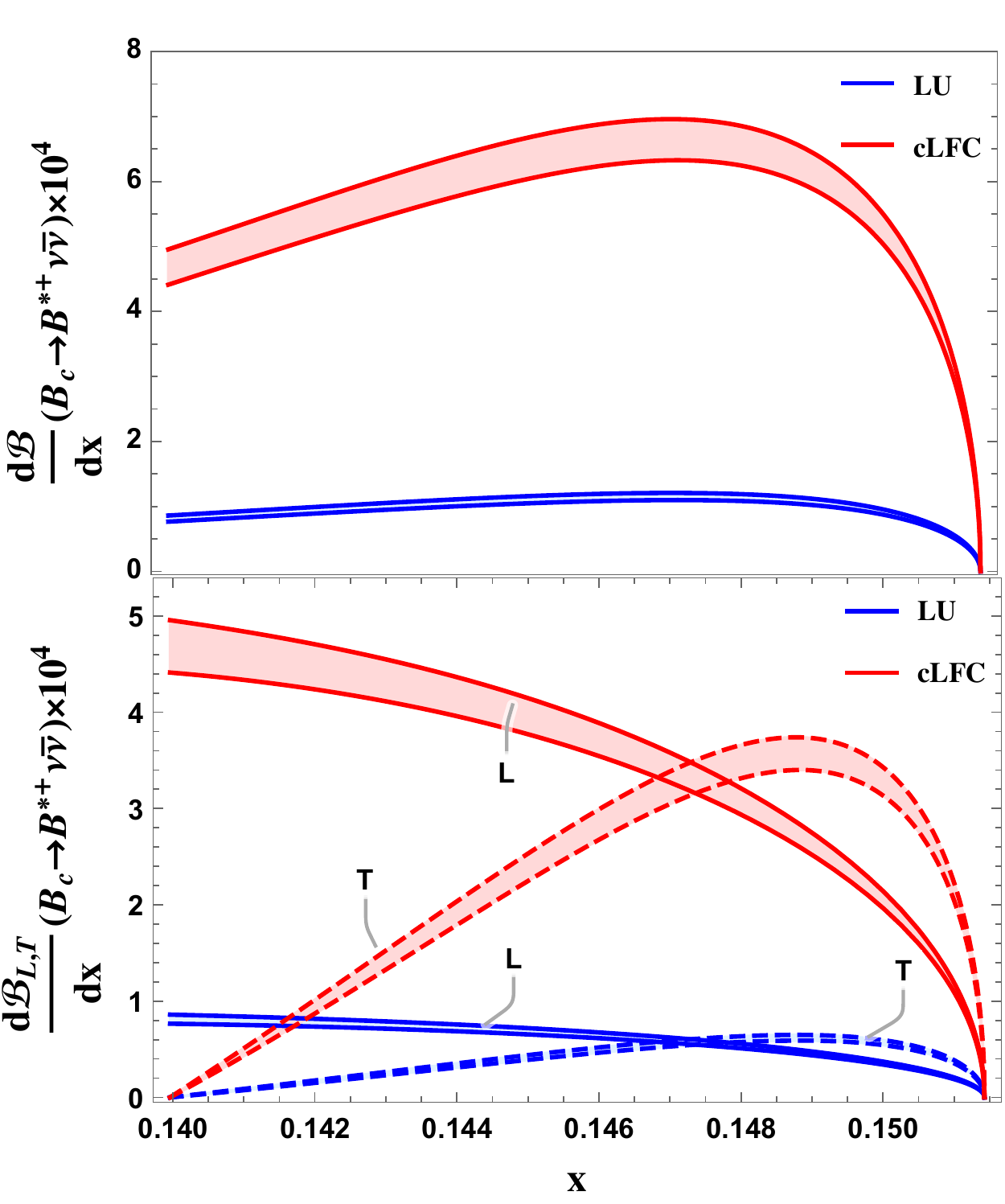}
    \caption{\small Missing energy distributions $d{\cal B}(B_c^+ \to B^{*+} \nu {\bar \nu})/dx$ (top) and $d{\cal B}_{L,T}(B_c^+ \to B^{*+} \nu {\bar \nu})/dx$ (bottom) for $x_U^{max}$ in  Eqs.~(\ref{xULU}) (blue curves) and  (\ref{xUcLFC}) (red curves). The widths of the curves  are obtained varying  the  form factor parameters.}\label{dBdxBustar}
\end{center}
\end{figure}

 As inferred from \eqref{dGBu} and \eqref{BustarT},  ${\cal B}(B_c^+ \to B^+ \nu {\bar \nu})$ depends on the combination $x_U^+$  in   Eq.~\eqref{xUpiu}, 
${\cal B}_L(B_c^+ \to B^{*+} \nu {\bar \nu})$   depends on $x_U^-$, and
${\cal B}_T(B_c^+ \to B^{*+} \nu {\bar \nu})={\cal B}_+(B_c^+ \to B^{*+} \nu {\bar \nu})+{\cal B}_-(B_c^+ \to B^{*+} \nu {\bar \nu})$ depends on both combinations.
 \begin{table}[!tb]
\center{\begin{tabular}{|l|l|}
\hline
\multicolumn{2}{|c|}{
{\rm Constants and quark masses}}
\\
\hline
$G_F = 1.16637(1)\times 10^{-5}\gev^{-2}$ \hfill  \cite{Zyla:2020zbs}& $m_c(m_c) = 1.279(8) \gev$ \hfill\cite{Chetyrkin:2017lif} \\
$M_W = 80.385(15) \gev$\hfill\cite{Zyla:2020zbs}   & $m_b(m_b)=4.163(16)\gev$\hfill\cite{Chetyrkin:2009fv,Zyla:2020zbs}  \\
$\sin^2\theta_W = 0.23121(4)$\hfill\cite{Zyla:2020zbs}  & $m_t(m_t) = 162.5\pm^{2.1}_{1.5}\gev$\hfill\cite{Zyla:2020zbs}\\
$\alpha(M_Z) = 1/127.9$\hfill\cite{Zyla:2020zbs}& $M_t=172.76(30) \gev$\hfill\cite{Zyla:2020zbs} 	
 \\
 $\alpha_s^{(5)}(M_Z)= 0.1179(10) $\hfill\cite{Zyla:2020zbs}  & \\
\hline
\multicolumn{2}{|c|}{
{\rm Meson masses and lifetimes}}
\\
\hline
$m_{K^+}= 493.677(13)\mev$	\hfill\cite{Zyla:2020zbs} & 
$\tau(K^+)= 1.2380(20)\times 10^{-8}\,\text{s}$\hfill\cite{Zyla:2020zbs}
\\
$m_{K^0}= 497.611(13)\mev$	\hfill\cite{Zyla:2020zbs} & $\tau(K_S)= 0.8954(4)\times 10^{-10}\,\text{s}$\hfill\cite{Zyla:2020zbs}	 \\
& $\tau(K_L)= 5.116(21)\times 10^{-8}\,\text{s}$\hfill\cite{Zyla:2020zbs} \\
$m_{B_d}= 5279.63(20)\mev$\hfill\cite{Zyla:2020zbs} & $\tau(B_d)= 1.519(4) \,\text{ps}$\hfill\cite{Zyla:2020zbs} \\
$m_{B^+}= 5279.25(26)\mev$\hfill\cite{Zyla:2020zbs} & \\
$m_{B^{*+}}= 5324.70(21)\mev$\hfill\cite{Zyla:2020zbs} & \\
$m_{B_s} =5366.88(14)\mev$\hfill\cite{Zyla:2020zbs} &
$\tau(B_s)= 1.515(4)\,\text{ps}$\hfill\cite{Zyla:2020zbs}\\
$m_{B_c} =6274.9(8)\mev$\hfill\cite{Zyla:2020zbs} &
$\tau(B_c)= 0.510(9)\,\text{ps}$\hfill\cite{Zyla:2020zbs}\\
\hline
\multicolumn{2}{|c|}{
{\rm Decay constants and parameters related to $\Delta F=2$ observables}}
\\
\hline
$F_K = 156.1(11)\mev$\hfill\cite{Aoki:2019cca}												& $\hat B_K= 0.7625(97)$\hfill\cite{Aoki:2019cca} \\
$\Delta M_K= 0.5293(9)\times 10^{-2} \,\text{ps}^{-1}$\hfill\cite{Zyla:2020zbs}	&
$|\epsilon_K|= 2.228(11)\times 10^{-3}$\hfill\cite{Zyla:2020zbs}\\
$F_{B_d} =190.0(1.3)\mev$\hfill\cite{Aoki:2019cca} & 
$F_{B_d} \sqrt{\hat B_{B_d}} = 216(10)\mev$\hfill\cite{Aoki:2019cca} \\
$F_{B_s} = 230.3(1.3)\mev$\hfill\cite{Aoki:2019cca} & 
$F_{B_s} \sqrt{\hat B_{B_s}} =262(10)\mev$\hfill\cite{Aoki:2019cca}  \\
$\eta_B=0.55(1)$\hfill\cite{Buras:1990fn,Urban:1997gw} &   \\
$\Delta M_d = 0.5065(19) \,\text{ps}^{-1}$\hfill\cite{Zyla:2020zbs} & 
$S_{J/\psi K_S}= 0.695(19)$\hfill\cite{Zyla:2020zbs}\\
$\Delta M_s = 17.756(21)\,\text{ps}^{-1}$\hfill\cite{Zyla:2020zbs} &
$S_{J/\psi\phi}= 0.054(20)$\hfill\cite{Aoki:2019cca}\\
\hline
\multicolumn{2}{|c|}{
{\rm CKM parameters}}
\\
\hline
$|V_{us}|=0.2252(5)$\hfill\cite{Zyla:2020zbs} & $|V_{cb}|=(41.0\pm1.4)\times
10^{-3}$\hfill\cite{Zyla:2020zbs}\\
$|V_{ub}|=3.72\times10^{-3}$\hfill\cite{Zyla:2020zbs} & $\gamma=68^\circ$\hfill\cite{Zyla:2020zbs}
\\
\hline
$ |V_{cd}|=0.22507$ & $|V_{cs}|=0.97348$\\
$ |V_{td}|=0.00856$ & $|V_{ts}|=0.04027$\\
 \hline
\end{tabular}  }
\caption {Parameters used in the analysis.}
\label{tab:par}
\end{table}
Using the parameters in Table~\ref{tab:par} and the  central values  for the form factors \cite{Cooper:2020wnj,Colangelo:2021dnv}
 we obtain:
\be
 {\cal B}(B_c^+ \to B^+ \nu {\bar \nu}) =  7.8 \times 10^8 |C_L+C_R|^2  = 6.9 \times 10^{-9}\, x_U^+ \label{BuxUpiu} 
 \ee
 and
 \bea
{\cal B}_L(B_c^+ \to B^{*+} \nu {\bar \nu})&  =& 1.2 \times 10^9 |C_L-C_R|^2  = 1.0 \times 10^{-8} x_U^- \nn \\
{\cal B}_T(B_c^+ \to B^{*+} \nu {\bar \nu})& = & 1.9 \times 10^7 |C_L+C_R|^2 +7.4 \times 10^8 |C_L-C_R|^2 \nn \\ & = & 1.7 \times 10^{-10} x_U^+ +6.5 \times 10^{-9} x_U^- \label{BustarxUmeno}  \\
{\cal B}(B_c^+ \to B^{*+} \nu {\bar \nu})& = & 1.9 \times 10^7 |C_L+C_R|^2 +1.9 \times 10^9 |C_L-C_R|^2\nn \\ &  = & 1.7 \times 10^{-10} x_U^+ +1.7 \times 10^{-8} x_U^- . \nn
\eea
The largest values of ${\cal B}(B_c^+ \to B^+ \nu {\bar \nu})$  and ${\cal B}(B_c^+ \to B^{*+} \nu {\bar \nu})$ correspond to the  largest  $x_U^+$ and $x_U^-$, respectively.
We scan  the branching fractions using $x_U^-=2 \, x_U-x_U^+$ and varying $0\le x_U^+ \le 2 \, x_U$, with $x_U\le x_U^{\max}$ for the two cases   \eqref{xULU}  (LU bound).  We extend the computation up to the  cLFC bound \eqref{xUcLFC}, which has been established by the analysis of the charged lepton modes, to investigate the size of the enhancement in this case.
In Fig.~\ref{dBdxBu} we show the largest  enhancement for the $d{\cal B}(B_c^+ \to B^+ \nu {\bar \nu})/dx$ distribution  obtained for  $x_U^+=2 \,x_U$. 
%
In Fig.~\ref{dBdxBustar}  we depict 
 the maximum enhancement for the  missing energy distribution and for the distributions of longitudinally and transversely polarized $B^{*+}$  in  $B_c^+ \to B^{*+} \nu {\bar \nu}$.
%
Integrating over $x$ we have:
\bea
{\cal B}(B_c^+ \to B^+ \nu {\bar \nu})_{\rm LU}^{\rm max}&=&(4.7 \pm 0.25 ) \times 10^{-7}\nn\\
{\cal B}(B_c^+ \to B^+ \nu {\bar \nu})_{\rm cLFC}^{\rm max}&=&(2.7 \pm 0.15 ) \times 10^{-6} \label{res1}\\
{\cal B}(B_c^+ \to B^{*+} \nu {\bar \nu})_{\rm LU}^{\rm max}&=&(1.1 \pm 0.06 ) \times 10^{-6}\nn\\
{\cal B}(B_c^+ \to B^{*+} \nu {\bar \nu})_{\rm cLFC}^{\rm max}&=&(6.5 \pm 0.3 ) \times 10^{-6} . \label{res2}
\eea
\begin{figure}[!tb]
\begin{center}
\includegraphics[width = 0.5\textwidth]{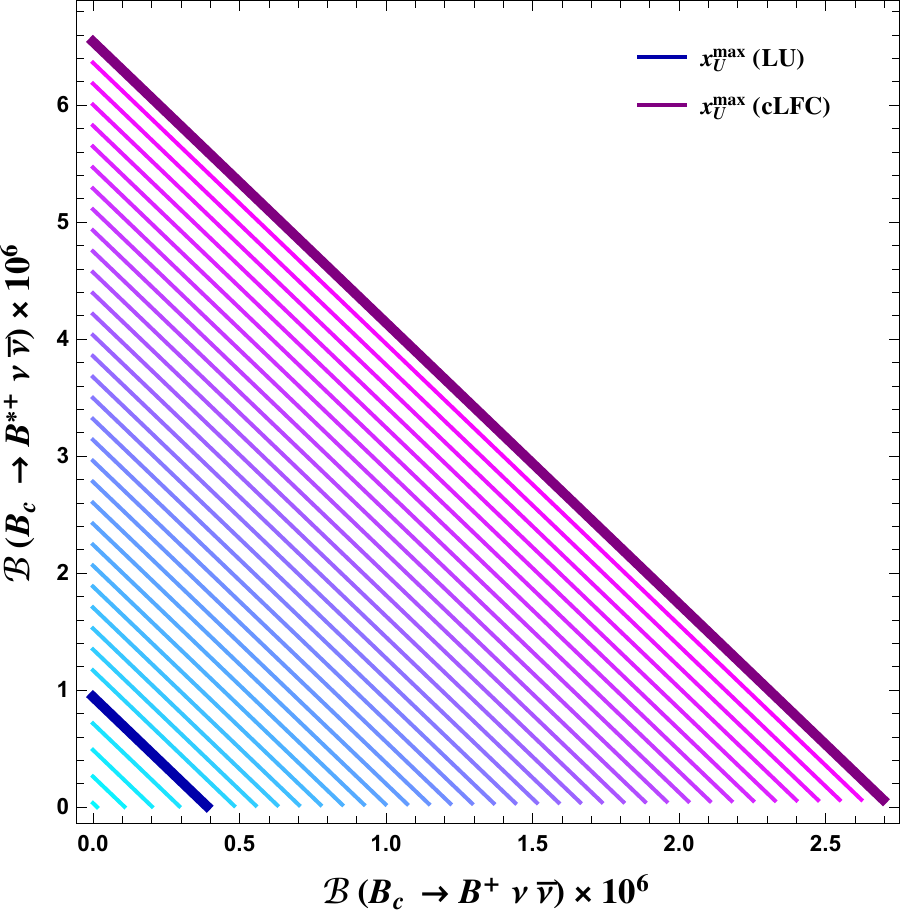}
    \caption{\small Anticorrelation between ${\cal B}(B_c^+ \to B^+ \nu {\bar \nu})$ and ${\cal B}(B_c^+ \to B^{*+} \nu {\bar \nu})$, varying  the combination of the coefficients  $x_U$  up to the bounds in Eqs.~(\ref{xULU}) and (\ref{xUcLFC}). Colors   from cyan to magenta indicate increasing values of $x_U$. The dark  blue line corresponds to the value saturating the LU bound and  the dark purple line to the value saturating the cLFC bound.  The colors corresponding to three other representative values of $x_U$ ($2,59,112$) are also indicated in the legend.}\label{correlation}
\end{center}
\end{figure}
The largest values of the branching fractions must be compared with the SM prediction  from  Eq.~\eqref{eq:CLSM}:
${\cal B}(B_c^+ \to B^+ \nu {\bar \nu})^{\rm SM}=(8.5 \pm 0.5) \times 10^{-18}$, 
${\cal B}(B_c^+ \to B^{*+} \nu {\bar \nu})^{\rm SM}=(2.1 \pm 0.1) \times 10^{-17}$, and with the estimate of the long-distance contributions  discussed in the Appendix. Hence,  
a  huge enhancement with respect to tiny SM prediction is possible. 
Setting $x_U$ below the bounds  (\ref{xULU}) and (\ref{xUcLFC})  and varying $x_U^+ \in[0,\,2 \, x_U]$ the branching fractions can be read in the plot in Fig.~\ref{correlation}. 

The enhancements  in Eqs.~\eqref{res1} and \eqref{res2}, achievable in  generic NP scenarios, must be taken with caution, since  they would be the manifestation of BSM phenomena   affecting  other  processes to a level that is necessary to control. For this reason it is worth  considering a well defined extension of the Standard Model, as discussed  in the next section.

\section{$c \to u \nu {\bar \nu}$ transition in the 331 model}\label{sec:331}
 Among the extensions of the Standard Model   we focus on the 331 models,  a class of  models based on the gauge group $SU(3)_C \times SU(3)_L \times U(1)_X$ \cite{Pisano:1991ee,Frampton:1992wt}. The gauge symmetry  is spontaneously broken to the SM group    $SU(3)_C \times SU(2)_L \times U(1)_Y$, followed by the  spontaneous breaking to $SU(3)_C \times U(1)_Q$. 
 This extension of the gauge group  has remarkable features.
 Left-handed fermions transform under $SU(3)_L$ either as triplets or as antitriplets.
The requirement of gauge anomaly cancellation imposes that the number of triplets should be equal to the number of antitriplets. This constraint together  with  the asymptotic freedom of QCD imposes that the number of fermion generations is equal to the number of colors,  a hint of why there are three generations in nature.
The quark generations transform differently under  $SU(3)_L$, and a possibility is that   two left-handed quark generations  transform as  $SU(3)_L$ triplets and one as an antitriplet. Choosing the latter one as the third generation, the different assignment can be a hint of why there is a large top mass.

The electric charge generator $Q$ is defined by
 \be
 Q=T_3+\beta T_8+X \,\, , \label{eq:defbeta}
 \ee
 with $T_3$ and $T_8$ the diagonal $SU(3)_L$  and $X$  the  $U(1)_X$ generators. 
 The parameter $\beta$ defines the  specific variant of the model. Four new gauge bosons  have integer electric charges if $\beta$  is a multiple of $\displaystyle{\frac{1}{\sqrt{3}}} $ and  $\sqrt{3}$. 
The $U(1)_X$ gauge coupling $g_X$ and the $SU(3)_L$ coupling  $g$ are related,
\be
{g_X^2 \over g^2}={6 \sin^2 \theta_W \over 1-(1+\beta^2) \sin^2 \theta_W}\,\,. \label{gXg}
\ee
  Equation \eqref{gXg} provides the bound $\dd |\beta| \le \frac{1}{\tan \theta_W (M_{Z^\prime})}$ which  corresponds to $|\beta|<1.737$ for  
 the sine of the Weinberg angle
  $\sin \theta_W (M_{Z^\prime} = 1 \, {\rm TeV}) = 0.249$.

 In all the 331 variants there is  a  neutral gauge boson $Z^\prime$ mediating tree-level FCNC  in the quark sector, with universal and diagonal $Z^\prime$ couplings to leptons. 
The extended Higgs sector involves three $SU(3)_L$ triplets and one sextet.
New heavy fermions  are also present in the spectrum.

As in  the SM, quark mass eigenstates are defined upon rotation of flavor eigenstates through two unitary matrices, $U_L$ (for up-type quarks) and $V_L$ (for down-type quarks). The relation $V_{CKM}=U_L^\dagger V_L$ holds. However, while in  the SM $V_{CKM}$ only enters  in charged current interactions  and the two rotation matrices do not appear individually.  In 331 model only one matrix,  either $U_L$ or $V_L$, can be expressed in terms of $V_{CKM}$ and of the other one. The remaining rotation matrix affects the $Z^\prime$ couplings to the quarks.  Choosing  $V_L$ as the surviving rotation matrix,  it can be  parametrized  as
 \begin{equation}
V_L=\left(\begin{array}{ccc}
{\tilde c}_{12}{\tilde c}_{13} & {\tilde s}_{12}{\tilde c}_{23} e^{i \delta_3}-{\tilde c}_{12} {\tilde s}_{13} {\tilde s}_{23}e^{i(\delta_1
-\delta_2)} & {\tilde c}_{12}{\tilde c}_{23} {\tilde s}_{13} e^{i \delta_1}+ {\tilde s}_{12} {\tilde s}_{23}e^{i(\delta_2+\delta_3)} \\
-{\tilde c}_{13} {\tilde s}_{12}e^{-i\delta_3} & {\tilde c}_{12}{\tilde c}_{23} + {\tilde s}_{12}
 {\tilde s}_{13} {\tilde s}_{23}e^{i(\delta_1-\delta_2-\delta_3)} & - {\tilde s}_{12} {\tilde s}_{13}{\tilde c}_{23}e^{i(\delta_1 -\delta_3)}
-{\tilde c}_{12} {\tilde s}_{23} e^{i \delta_2} \\
- {\tilde s}_{13}e^{-i\delta_1} & -{\tilde c}_{13} {\tilde s}_{23}e^{-i\delta_2} & {\tilde c}_{13}{\tilde c}_{23}
\end{array}\right) \,\,\,\,\label{VL-param}
\end{equation}
with $\tilde c_i=\cos \theta_i$, $\tilde s_i=\sin \theta_i$, and  phases $\delta_{1,2,3}$.
With this parametrization, considering  the  $Z^\prime$ couplings to the quarks, one finds
that the $B_d$ system involves the parameters ${\tilde s}_{13}$ and $\delta_1$,   the $B_s$ system 
${\tilde s}_{23}$ and $\delta_2$, and  the kaon system ${\tilde s}_{13}$, ${\tilde s}_{23}$ and $\delta_2 - \delta_1$. This  provides remarkable correlations among observables in kaon,  $B_{d,}$ and  $B_s$  systems \cite{Buras:2012dp,Buras:2013dea,Buras:2014yna,Buras:2015kwd,Buras:2016dxz}.

It is interesting to observe that  the relation 
\be
U_L=V_L \cdot V_{CKM}^\dagger\,
\label{UL}
\ee
allows to bound the $Z^\prime$ mediated FCNC  transitions of up-type quarks  using the constraints established in the down-type quark sector   \cite{Buras:2021rdg}. Such a relation connecting the down-type and up-type quark  FCNC processes is a peculiar feature of the 331  model. 

The $Z^\prime$ coupling to ordinary fermions, for a generic value of the $\beta$ parameter, is encoded in the 331 Lagrangian density:

\bea\label{ZprimeFR}
i\,L_{int}^{Z^\prime} &=&  i{g  Z^{\prime \mu}\over 2 \sqrt{3} c_W \sqrt{1-(1+\beta^2) s_W^2}} \nn\\
 && \Bigg\{ \sum_{\ell=e,\mu,\tau}\Big\{\left[1-(1+\sqrt{3} \beta) s_W^2 \right] \left({\bar \nu}_{\ell \, L} \gamma_\mu \nu_{\ell \, L}+ {\bar \ell}_L \gamma_\mu \ell_L \right)- 2 \sqrt{3} \beta s_W^2 {\bar \ell}_R \gamma_\mu \ell_R 
\Big\}  \nn \\
&+&
\sum_{i,j=1,2,3} \Big\{ \big[-1+(1+{\beta \over \sqrt{3}})s_W^2\big]({\bar q}_{uL})_i \gamma_\mu (q_{uL})_j \delta_{ij}+ 2c_W^2 ({\bar q}_{uL})_i \gamma_\mu (q_{uL})_j u_{3i}^* u_{3j} \nn \\
&+& \big[-1+(1+{\beta \over \sqrt{3}})s_W^2\big]({\bar q}_{dL})_i \gamma_\mu (q_{dL})_j \delta_{ij}+ 2c_W^2 ({\bar q}_{dL})_i \gamma_\mu (q_{dL})_j v_{3i}^* v_{3j} \nn \\
&+&{4 \over \sqrt{3}}\beta s_W^2({\bar q}_{uR})_i \gamma_\mu (q_{uR})_j \delta_{ij}
-{2 \over \sqrt{3}}\beta s_W^2({\bar q}_{dR})_i \gamma_\mu (q_{dR})_j \delta_{ij}\Big\}
\,\,,
\eea
where $s_W=\sin \theta_W, c_W=\cos \theta_W$, $q_u \, (q_d)$ denotes an up (down)-type quark ($i,\,j$ are generation indices), and $v_{ij}$ and $u_{ij}$ are the elements of  the $V_L$ and $U_L$ matrices, respectively.
 The models corresponding to $\beta=\pm\displaystyle\frac{2}{\sqrt{3}}$ and $\beta=\pm\displaystyle\frac{1}{\sqrt{3}}$, together with  the choice of  the fermions in the third generation as transforming as $SU(3)_L$ antitriplets, satisfy a number of phenomenological constraints  \cite{Buras:2013dea}.  In particular, it is possible to select a region of the parameter space compatible with the constrains from $\Delta F=2$ observables in the $B_d,\,B_s$ and  $K$ systems and from the  electroweak precision observables,  provided that the $Z^\prime$ mass is not lighter  than  $1$ TeV. 
In the variant with $\beta=\displaystyle\frac{2}{\sqrt{3}}$ relevant contributions are 
 predicted to the ratio $\displaystyle\frac{\varepsilon^\prime}{\varepsilon}$  \cite{Buras:2015kwd}. 

As shown in \cite{Buras:2014yna}, the $Z-Z^\prime$ mixing  can be neglected in $\Delta F=2$ transitions, while it must be taken into account in decays with neutrinos in the final state.  The $Z-Z^\prime$  mixing angle is written as \cite{Buras:2014yna}
\be\label{sxi}
\sin\xi=\frac{c_W^2}{3} \sqrt{f(\beta)}\left(3\beta \frac{s_W^2}{c_W^2}+\sqrt{3}a\right) \frac{M_Z^2}{M_{Z^\prime}^2}
= B(\beta,a) \frac{M_Z^2}{M_{Z^\prime}^2} \,\, ,
\ee
where 
\be\label{central}
f(\beta)=\frac{1}{1-(1+\beta^2)s_W^2}  > 0\, 
\ee
and 
\be\label{ratiov}
-1 < a=\frac{v_-^2}{v_+^2}< 1 \,\, . 
\ee
 $v^2_\pm$  are given in terms of the 
vacuum expectation values of two  Higgs triplets $\rho$ and $\eta$:
\be
v_+^2=v_\eta^2+v_\rho^2\,\, , \qquad \qquad  v_-^2=v_\eta^2-v_\rho^2\,\, .
\ee
The parameter $a$  is expressed in terms of  $\dd \tan\bar\beta=\frac{v_\rho}{v_\eta}$  as in two Higgs doublet models (we use    $\bar \beta$  to distinguish this parameter from  $\beta$  defining the 331 model  in \eqref{eq:defbeta}) \cite{Buras:2014yna}:
\be\label{basica}
a=\frac{1-\tan^2\bar\beta}{1+\tan^2\bar\beta}\,\,.
\ee

We consider the four variants  scrutinized in \cite{Buras:2013dea}.
For the  modes with a neutrino-antineutrino pair  in the final state, the $Z- Z^\prime$ mixing is included   replacing 
\be
 \Delta_L^{\nu {\bar \nu}}(Z^\prime) \to \Delta_L^{\nu {\bar \nu}}(Z^\prime)(1+R_{\nu \bar \nu}^L(a)) .
 \ee
$R_{\nu \bar \nu}^L(a)$ is defined as
\be R_{\nu \bar \nu}^L(a)=B(\beta,a)\displaystyle \frac{\Delta_{\nu {\bar \nu}}(Z)}{\Delta_L^{\nu {\bar \nu}}(Z^\prime)} \,\, , \label{Rnunu}
\ee  
with $B(\beta,a)$ in \eqref{sxi}
and $\Delta_{\nu {\bar \nu}}(Z)$ the SM $Z$ coupling to neutrinos.

In the 331 model the $B_c \to B^{(*)+} \nu {\bar \nu}$ modes   present several  features.  The structure of the 331 model allows us to use data from $B$ and $K$ decays to constrain $c \to u $ modes. 
Moreover,  $Z^\prime$ mediates FCNC  at tree level only in the case of left-handed quarks; hence, the coefficient $C_R$ in the Hamiltonian \eqref{hamil} vanishes in all the model variants. 
\begin{figure}[!tb]
\begin{center}
\includegraphics[width = 0.3\textwidth]{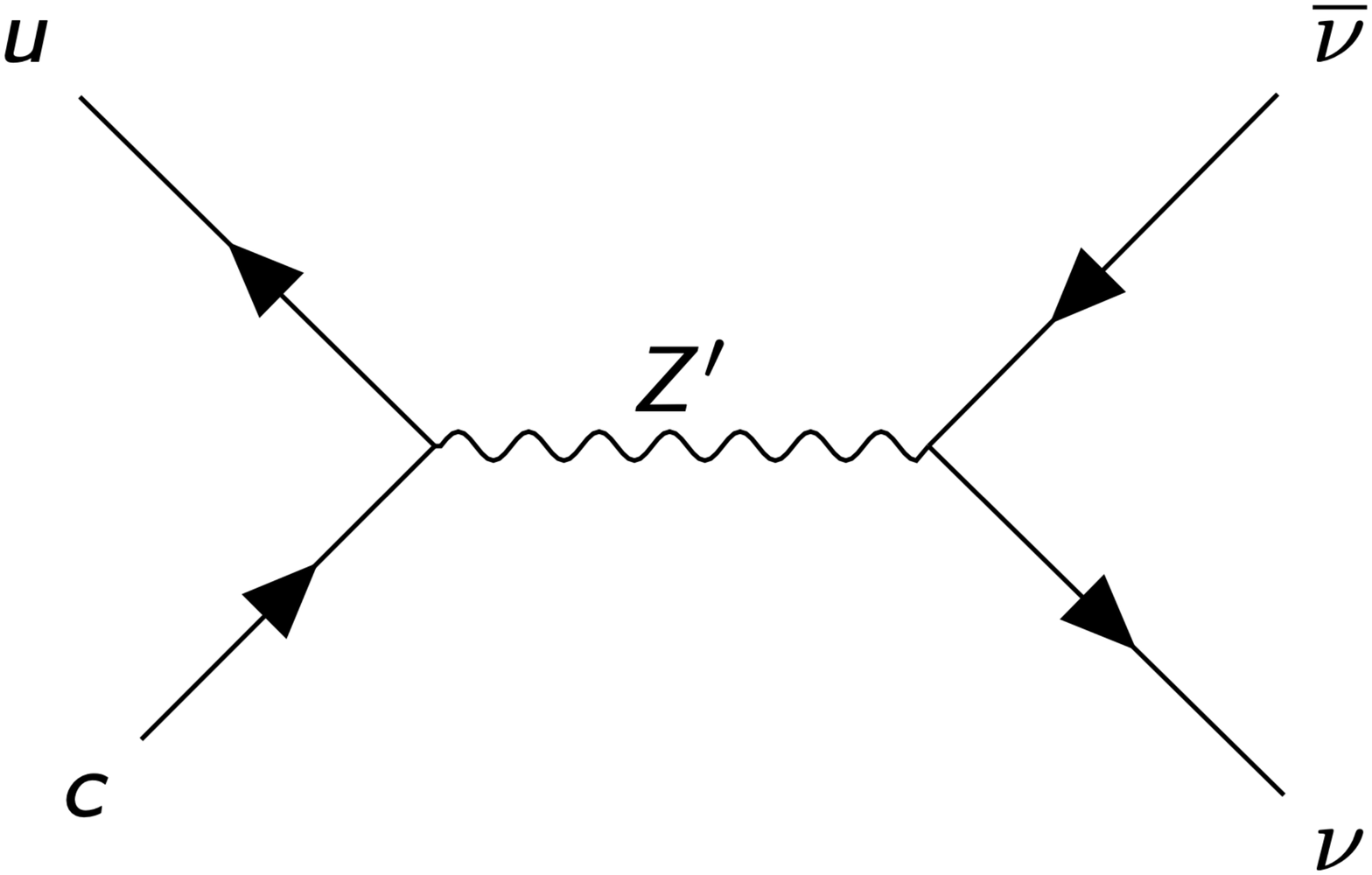}
    \caption{\small Tree-level $Z^\prime $ contribution to the  $c \to u \nu {\bar \nu}$ effective Hamiltonian. }\label{331diagram}
\end{center}
\end{figure}

Considering the  contribution from the tree-level diagram in Fig.~\ref{331diagram} and 
using  the coupling of $Z^\prime$ to quarks   and neutrinos  derived from Eq.~(\ref{ZprimeFR}), the coefficient  $C_L$  in \eqref{hamil} reads:
\be
C_L^{331}=\frac{\Delta_L^{uc}(Z^\prime) \, \Delta_L^{\nu {\bar \nu}}(Z^\prime)}{M_{Z^\prime}^2}\big(1+R_{\nu \bar \nu}^L(a)\big) \, , \label{CL331}
\ee
where 
\bea
\Delta_L^{uc}(Z^\prime)&=&\frac{g \, c_W}{\sqrt{3} \, \sqrt{1-(1+\beta^2) s_W^2}}u_{31}^* u_{32}
\nn \\
\Delta_L^{\nu {\bar \nu}}(Z^\prime)&=&\frac{g~\left[1-(1+\sqrt{3}\beta)s_W^2\right]}{2 \sqrt{3}c_W\sqrt{1-(1+\beta^2)s_W^2}}\,\, .
\eea
The elements $u_{ij}$ are obtained from Eqs.~(\ref{VL-param}) and (\ref{UL}). As a consequence,
$C_L$ depends on the  parameters ${\tilde s}_{13},\,\delta_1$, 
 ${\tilde s}_{23},\, \delta_2$ that  in pair control the $B_d$ and $B_s$ decays, respectively, and altogether  govern the $K$ decays.
 $C_L$  also depends on the $Z-Z^\prime$ mixing parameter $a$. The $B_c^+ \to B^{(*)+} \nu {\bar \nu}$ processes must be studied in such a parameter space.
 
 \section{  $B_c^+ \to B^{(*)+} \nu {\bar \nu}$ decays in the 331 model}\label{sec:331modes}

In the numerical analysis of $B_c^+ \to B^{(*)+} \nu {\bar \nu}$ in the 331 model we follow the method described in  \cite{Buras:2015kwd}. We select the model parameters imposing that $\Delta M_{B_d}$, $S_{J/\psi K_S}$ and $\Delta M_{B_s}$, $S_{J/\psi \phi}$, whose measurements are quoted in Table \ref{tab:par},  lie in their experimental ranges within  $2\sigma$. In the kaon sector we require  that $\epsK$ is in the range  $[1.6,\,2.5] \times 10^{-3}$ and $\Delta M_K$ varies between $[0.75,\,1.25] \times \left( \Delta M_K \right)_{SM}$, i.e., $\left(\Delta M_K\right)_{SM}=0.0047$ GeV  using $V_{ub}$ in Table \ref{tab:par}.
The  formulas for such observables in the SM and in 331 models  can be found in \cite{Buras:2012dp}. For $\epsK$ we use the updated result in \cite{Brod:2019rzc}. The other input quantities are also collected in Table \ref{tab:par}. For the  CKM matrix elements  the Table displays the four entries chosen as the independent ones,  the others  are  derived.

\begin{figure}[!tb]
\begin{center}
\includegraphics[width = 0.87\textwidth]{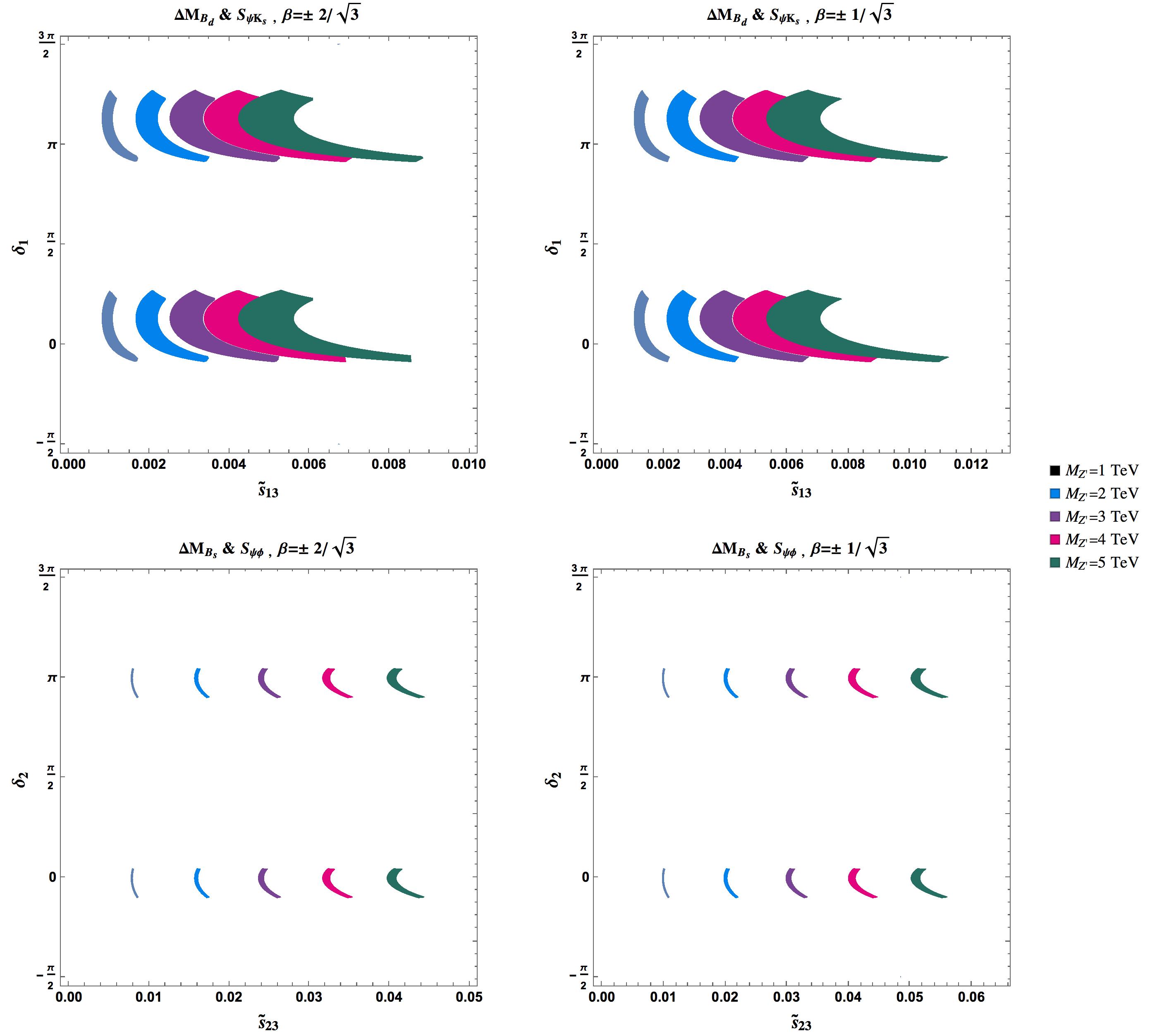}
    \caption{\small Allowed regions in the 331 space of parameters ${\tilde s}_{13},\,\delta_1$ (top) and
 ${\tilde s}_{23},\, \delta_2$ (bottom) for  $\beta=\pm \frac{2}{\sqrt{3}}$ (left) and $\beta= \pm \frac{1}{\sqrt{3}}$ (right),  varying the $Z^\prime$ mass in the range $[1,\,5]$ TeV.}\label{oases}
\end{center}
\end{figure}

 The obtained allowed regions in the parameter space ${\tilde s}_{13},\,\delta_1$, 
 ${\tilde s}_{23},\, \delta_2$ are  in Fig.~\ref{oases} for $M_{Z^\prime}\in[ 1,\,  5]\,$ TeV.
The  regions  $({\tilde s}_{13},\,\delta_1)$ are obtained imposing the constraints on $\Delta M_{B_d},\, S_{J/\psi K_S}$, the regions  $({\tilde s}_{23},\, \delta_2)$ using $\Delta M_{B_s},\, S_{J/\psi \phi}$.
%
 %
In our computation of the observables in the 331 model, we vary ${\tilde s}_{13},\,\delta_1$, 
 ${\tilde s}_{23},\, \delta_2$  in their allowed ranges and select the values for which the constraints  from $\Delta F=2$ processes in the kaon sector are  also satisfied. 
For each value of $\beta$ Fig.~\ref{oases} shows the presence of two ranges for the phases $\delta_{1,2}$ which are independent of  $M_{Z^\prime}$. Choosing the 331 parameters in the allowed  ranges,
 the  coefficient $C_L^{331}$ can be computed and the $B_c \to B^{(*)+} \nu {\bar \nu}$ branching fractions can be predicted.

\begin{figure}[!tb]
\begin{center}
\includegraphics[width = 0.6\textwidth]{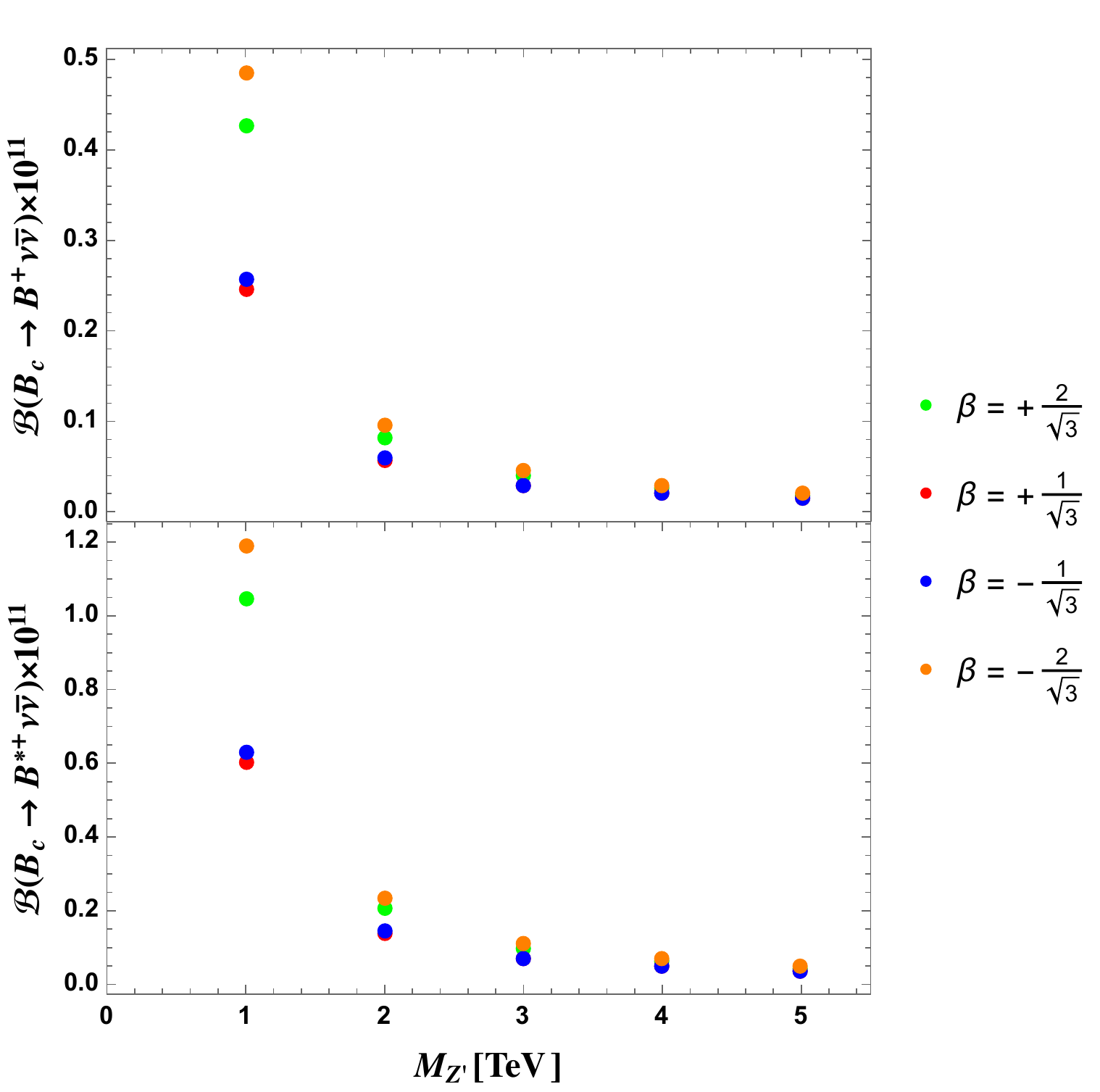}\\
    \caption{\small Branching ratios ${\cal B}(B_c^+ \to B^+ \nu {\bar \nu}) $ and ${\cal B}(B_c^+ \to B^{*+} \nu {\bar \nu})$ in 331 model for  $\beta=\pm \frac{2}{\sqrt{3}}$ and $\beta= \pm \frac{1}{\sqrt{3}}$, varying $M_{Z^\prime}$ from  1 to 5 TeV. The results correspond to the values of ${\tilde s}_{13},\,\delta_1$, 
 ${\tilde s}_{23},\, \delta_2$ and $a$ producing the largest rates. }\label{br331}
\end{center}
\end{figure}

In Fig.~\ref{br331} we plot  the missing energy distributions 
for the set of ${\tilde s}_{13},\,\delta_1$, 
 ${\tilde s}_{23},\, \delta_2$ and $a$  maximizing the  dineutrino  $B_c$ branching fractions  for each $M_Z^\prime$ up to 5 TeV. In all cases the choice $a=1$ provides the largest enhancement.
The central values of the branching fractions are:
\bea
{\cal B}(B_c^+ \to B^+ \nu {\bar \nu})&=& \frac{4.91}{\big(M_{Z^\prime}/{\rm GeV}\big)^4}  \hskip 1cm (\beta=-\frac{2}{\sqrt{3}}) \nn \\
{\cal B}(B_c^+ \to B^+ \nu {\bar \nu})&=& \frac{4.31}{\big(M_{Z^\prime}/{\rm GeV}\big)^4}  \hskip 1cm (\beta=+\frac{2}{\sqrt{3}}) \nn \\
{\cal B}(B_c^+ \to B^+ \nu {\bar \nu})&=& \frac{2.60}{\big(M_{Z^\prime}/{\rm GeV}\big)^4}    \hskip 1cm (\beta=-\frac{1}{\sqrt{3}}) \label{Bu331} \\
{\cal B}(B_c^+ \to B^+ \nu {\bar \nu})&=& \frac{2.48}{\big(M_{Z^\prime}/{\rm GeV}\big)^4}   \hskip 1cm (\beta=+\frac{1}{\sqrt{3}}) \nn 
\eea
and
\bea
{\cal B}(B_c^+ \to  B^{*+} \nu {\bar \nu})&=& \frac{12.02}{\big(M_{Z^\prime}/{\rm GeV}\big)^4}  \hskip 1cm (\beta=-\frac{2}{\sqrt{3}}) \nn \\
{\cal B}(B_c^+ \to B^{*+} \nu {\bar \nu})&=& \frac{10.56}{\big(M_{Z^\prime}/{\rm GeV}\big)^4}    \hskip 1cm (\beta=+\frac{2}{\sqrt{3}}) \nn \\
{\cal B}(B_c^+ \to B^{*+} \nu {\bar \nu})&=& \frac{6.36}{\big(M_{Z^\prime}/{\rm GeV}\big)^4}    \hskip 1cm (\beta=-\frac{1}{\sqrt{3}}) \label{Bus331} \\
{\cal B}(B_c^+ \to B^{*+} \nu {\bar \nu})&=& \frac{6.08}{\big(M_{Z^\prime}/{\rm GeV}\big)^4}   \hskip 1cm (\beta=+\frac{1}{\sqrt{3}}) . \nn \eea
The enhancement with respect to the SM is large, even though the branching fractions do not exceed  ${\cal O}(10^{-11})$. 

\section{Correlations between the modes $c \to u \nu {\bar \nu}$ and $s \to d  \nu {\bar \nu}$,   $b \to s  \nu {\bar \nu}$  in the 331 model}\label{sec:corr}
We have remarked that a peculiar feature of the 331 model is the possibility of constraining  FCNC up-type quark processes using  information on  FCNC  down-type quark transitions. On this basis we can establish  the correlations between   $B_c \to B^{(*)+} \nu {\bar \nu}$ and the  $s \to d \nu {\bar \nu}$ induced transitions $K^+ \to \pi^+ \nu {\bar \nu}$ and $K_L \to \pi^0 \nu {\bar \nu}$,  and  between $B_c \to B^{(*)+} \nu {\bar \nu}$ and $B  \to  \left\{ X_s,\,K, \, K^{*}\right\}\nu {\bar \nu}$ induced by $b \to s \nu {\bar \nu}$. 
In the SM such  transitions proceed through box and $Z^0$ penguin diagrams. 
The low-energy  $s \to d \nu {\bar \nu}$ Hamiltonian  reads in the SM:
\bea\label{Heffknunu}
H_{eff}^{{\bar s} \to {\bar d}\nu\bar\nu}\Big|_{SM}&=&4\frac{G_F}{\sqrt 2}\frac{\alpha}{2\pi\sin^2\theta_W}
\\
&&
 \sum_{\ell=e,\mu,\tau}{\left[V_{cs}^*V_{cd}
{X_\text{NNL}^\ell(x_c)}+V_{ts}^*V_{td} X(x_t)\right]}
(\bar s \gamma_\mu P_L d)(\bar\nu_\ell\gamma_\mu P_L \nu_\ell)+{\rm H.c.}\,, \nn
\eea
with $x_i=m_i^2/M_W^2$. {$X_\text{NNL}^\ell(x_c)$} takes into account the internal charm contribution \cite{Buchalla:1993wq,Buchalla:1998ba,Buras:2005gr,Buras:2006gb,Brod:2008ss,Brod:2010hi}, the function
\be
X(x_t)=\eta_X~{\frac{x_t}{8}}\;\left[{\frac{x_t+2}{x_t-1}}
+ {\frac{3 x_t-6}{(x_t -1)^2}}\; \ln x_t\right]
\ee
describes the internal top contribution. $\eta_X=0.994$ is a QCD correction  computed for 
  $m_t = m_t(m_t)$ \cite{Buchalla:1998ba,Misiak:1999yg}.
In the charged $K^+ \to \pi^+ \nu {\bar \nu}$ mode both  contributions must be taken into account,  in  $K_L \to \pi^0 \nu {\bar \nu}$   the top quark contribution dominates.
The top quark contribution also  dominates  in the  $b \to s \nu {\bar \nu}$ modes governed by the effective Hamiltonian 
\be
H_{eff}^{b \to s \nu {\bar\nu}}\Big|_{SM}=
4\frac{G_F}{\sqrt 2}\frac{\alpha}{2\pi\sin^2\theta_W}\sum_{\ell=e,\mu,\tau}
{\left[V_{ts}^* V_{tb} X(x_t)\right]}
(\bar s\gamma^\mu P_L b)(\bar\nu_\ell\gamma_\mu P_L\nu_\ell)
+{\rm H.c.}\,.
\ee
The 331  contribution from the tree-level $Z^\prime$ exchange can be  included in the  Hamiltonian replacing  $X(x_t) \to X(M)$ with
\be
X(M)=X(x_t)+\Delta X^i(M) \,\,  \label{XNP}
\ee
and
\bea\label{DeltaXKB}
\Bigg[4\frac{G_F}{\sqrt 2}\frac{\alpha}{2\pi\sin^2\theta_W} V_{ts}^*V_{td}\Bigg] \,\,\Delta X^{sd}(K)&=&\frac{\Delta_L^{sd}(Z')\Delta_L^{\nu\bar\nu}(Z')}{M_{Z'}^2} \label{XLK} \\
\Bigg[4\frac{G_F}{\sqrt 2}\frac{\alpha}{2\pi\sin^2\theta_W} V_{ts}^* V_{tb}\Bigg] \,\,\Delta X^{bs}(B)&=&\frac{\Delta_{L}^{bs}(Z')\Delta_{L}^{\nu\bar\nu}(Z')}{M_{Z'}^2} \,\, . \label{XLB}
\eea
The  $Z-Z^\prime$ mixing is included multiplying  the rhs of  Eqs.~(\ref{XLK}) and (\ref{XLB}) by  $(1+R_{\nu \bar \nu}^L(a))$, with $R_{\nu \bar \nu}^L$ in \eqref{Rnunu}.

\subsection{Correlations with dineutrino kaon modes}
In the SM the decays  $K^+ \to \pi^+ \nu {\bar \nu}$ and $K_L \to \pi^0 \nu {\bar \nu}$  are predicted with branching ratios  of ${\cal O}(10^{-11})$. The processes are  theoretically well controlled,  due to the possibility of relating their hadronic matrix elements to the precisely measured semileptonic  $K^+ \to \pi^0 e^+ \nu_e$ matrix element. 
  The NA62 Collaboration at CERN has measured ${\cal B}(K^+ \to \pi^+ \nu {\bar \nu})=(10.6 \pm^{4.0}_{ 3.4}|_{\rm stat} \pm 0.9_{\rm syst}) \times 10^{-11}$ at $68\%$ C.L.  \cite{NA62:2021zjw}. The upper bound for the neutral mode  is ${\cal B}(K_L \to \pi^0 \nu {\bar \nu})<3.9 \times 10^{-9}$ (at $90\%$ C.L)  \cite{Ahn:2018mvc,Ahn:2020opg}.
Detailed discussions of the dineutrino kaon  modes in the SM  and in the 331 model  are presented  in Refs.~\cite{Buras:2020xsm} and  \cite{Buras:2012dp,Buras:2013dea,Buras:2014yna,Buras:2015kwd,Buras:2016dxz}.
The  branching ratios are  expressed in the form

\bea
 {\cal B}(K^+\to \pi^+ \nu\bar\nu) &=& \kappa_+ (1+\Delta_{EM})\left [ \left ( \frac{{\rm Im} X_{\rm eff} }{\lambda^5}  \right )^2 + \left ( \frac{{\rm Re} X_{\rm eff} }{\lambda^5}
  - P_c(X)  \right )^2 \right ] , \label{BRSMKpiu} \\
{\cal B}( K_L \to \pi^0 \nu\bar\nu) &=& \kappa_L \left ( \frac{{\rm Im}  X_{\rm eff} }{\lambda^5} \right )^2 ,  \label{eq:BRSMKL} 
\eea
with $\lambda=|V_{us}|$ in Table \ref{tab:par}.  The   other quantities are
\bea
\kappa_+&=&(5.21\pm0.025) \left[\frac{\lambda}{0.2252}\right]^8 \times 10^{-11} \,\, , \nn \\
 \kappa_{\rm L}&=&(2.247\pm0.013) \left[\frac{\lambda}{0.2252}\right]^8 \times 10^{-10}  \,\, , \nn \\
P_c(X)&=&0.405\pm0.024   \,\, , \label{kappapiu} \\
\Delta_{EM}&=&-0.03 \,\, , \nn \\
X_{\rm eff} &= &V_{ts}^* V_{td} \, X(K) \,\, , \nn
\eea
with $X(K)$  in Eqs.~(\ref{XNP}) and (\ref{XLK}) \cite{Buras:2005gr,Buras:2006gb,Brod:2008ss,Isidori:2005xm,Mescia:2007kn}. \\

\begin{figure}[]
\begin{center}
\includegraphics[width = 0.8\textwidth]{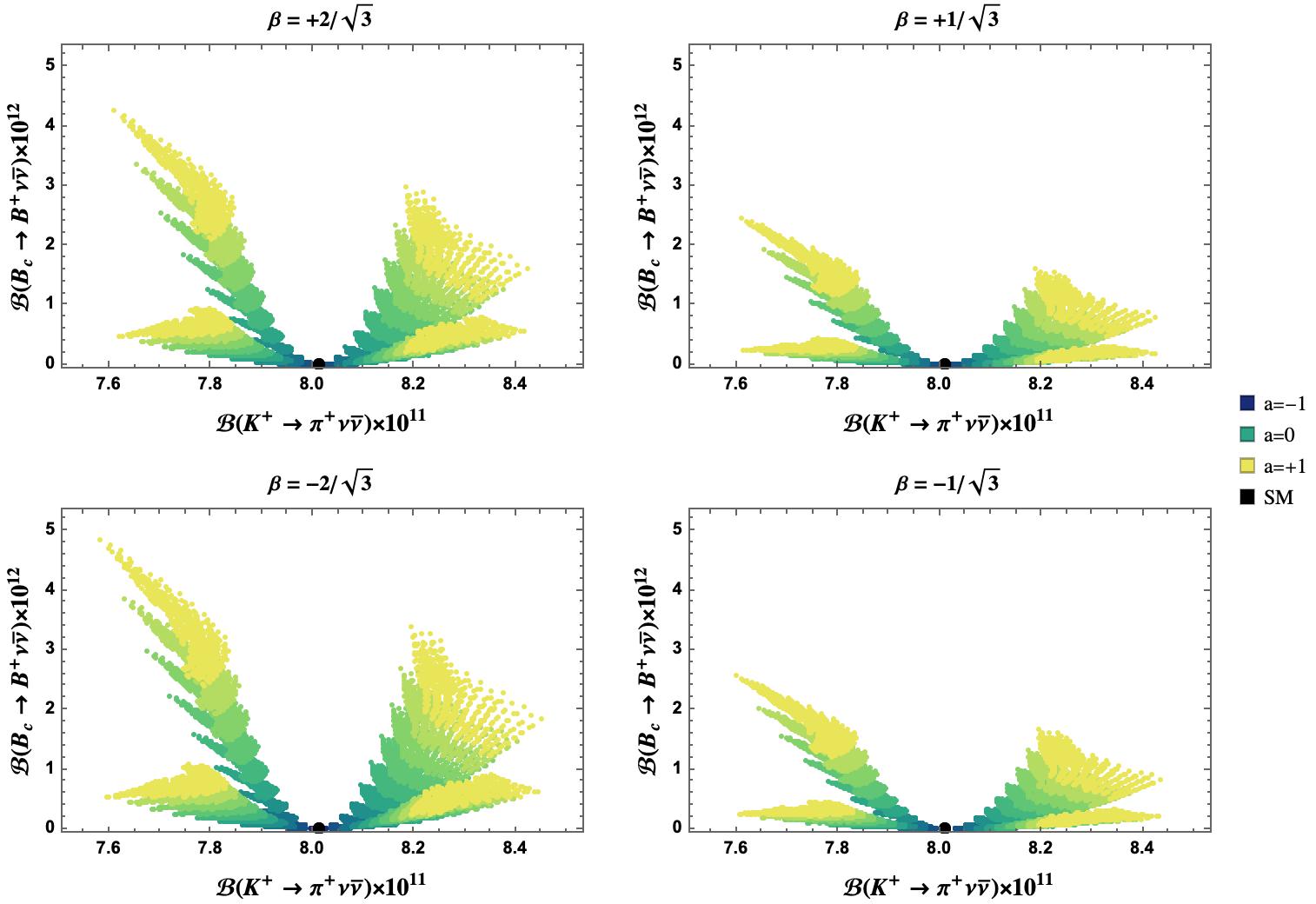}\\
    \caption{\small Correlations between  the branching fractions   ${\cal B}(B_c \to B^+ \nu \bar \nu)$ and ${\cal B}(K^+ \to \pi^+ \nu \bar \nu)$ in  the 331 model with $\beta= \frac{2}{\sqrt 3},  \frac{1}{\sqrt 3}, -\frac{2}{\sqrt 3}, -\frac{1}{\sqrt 3}$, $M_{Z^\prime}=1$ TeV and  $Z-Z^\prime$ mixing parameter $a=-1,0,1$. The black dot is the SM result.}\label{plKpiuBu}
\end{center}
\end{figure}

\begin{figure}[t]
\begin{center}
\includegraphics[width = 0.8\textwidth]{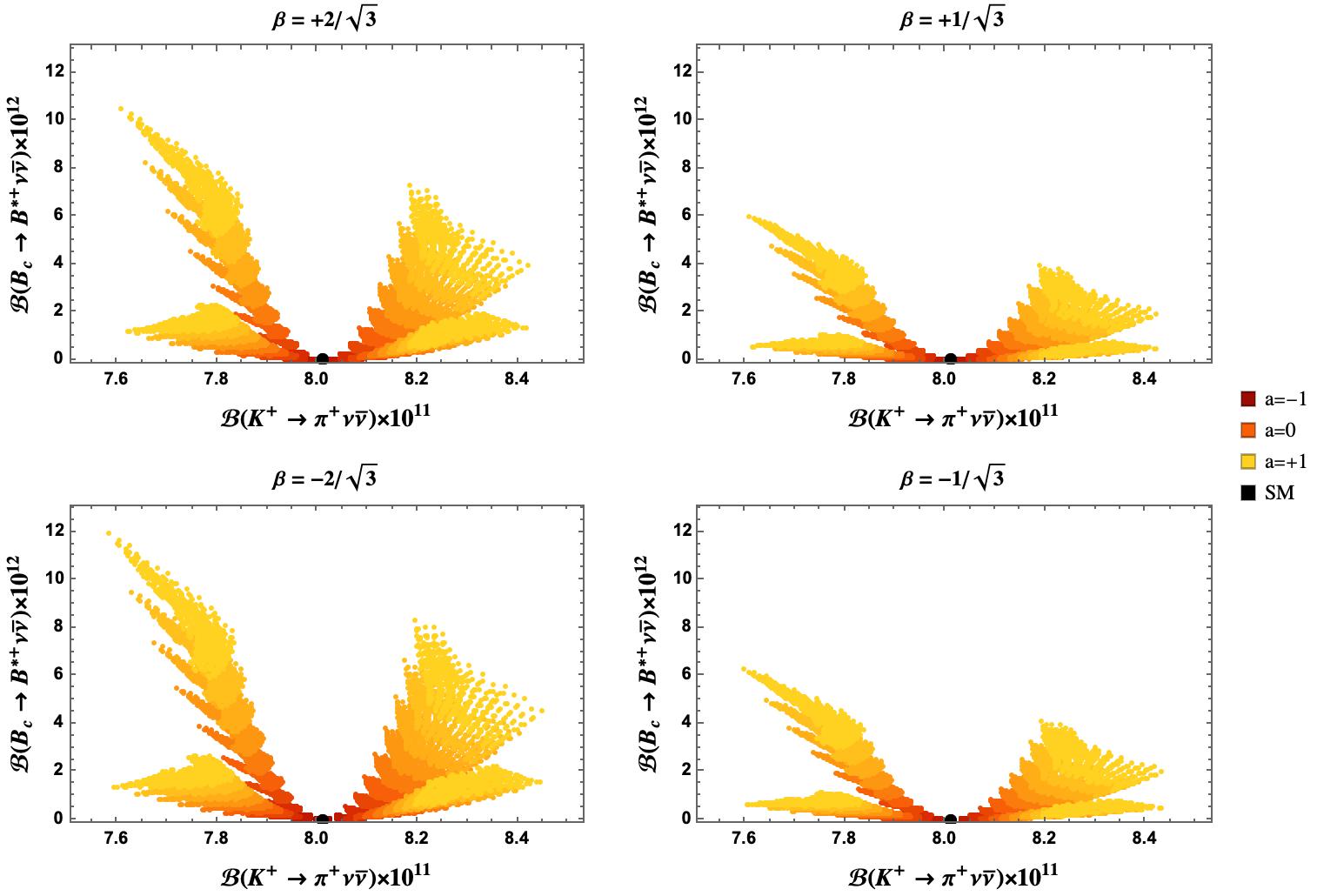}\\
    \caption{\small Correlations between   ${\cal B}(B_c \to B^{*+} \nu \bar \nu)$ and ${\cal B}(K^+ \to \pi^+ \nu \bar \nu)$ in the 331 model with parameters as in Fig.~\ref{plKpiuBu}. The black dot is the SM result.}
    \label{plKpiuBustar}
\end{center}
\end{figure}

\begin{figure}[]
\begin{center}
\includegraphics[width = 0.8\textwidth]{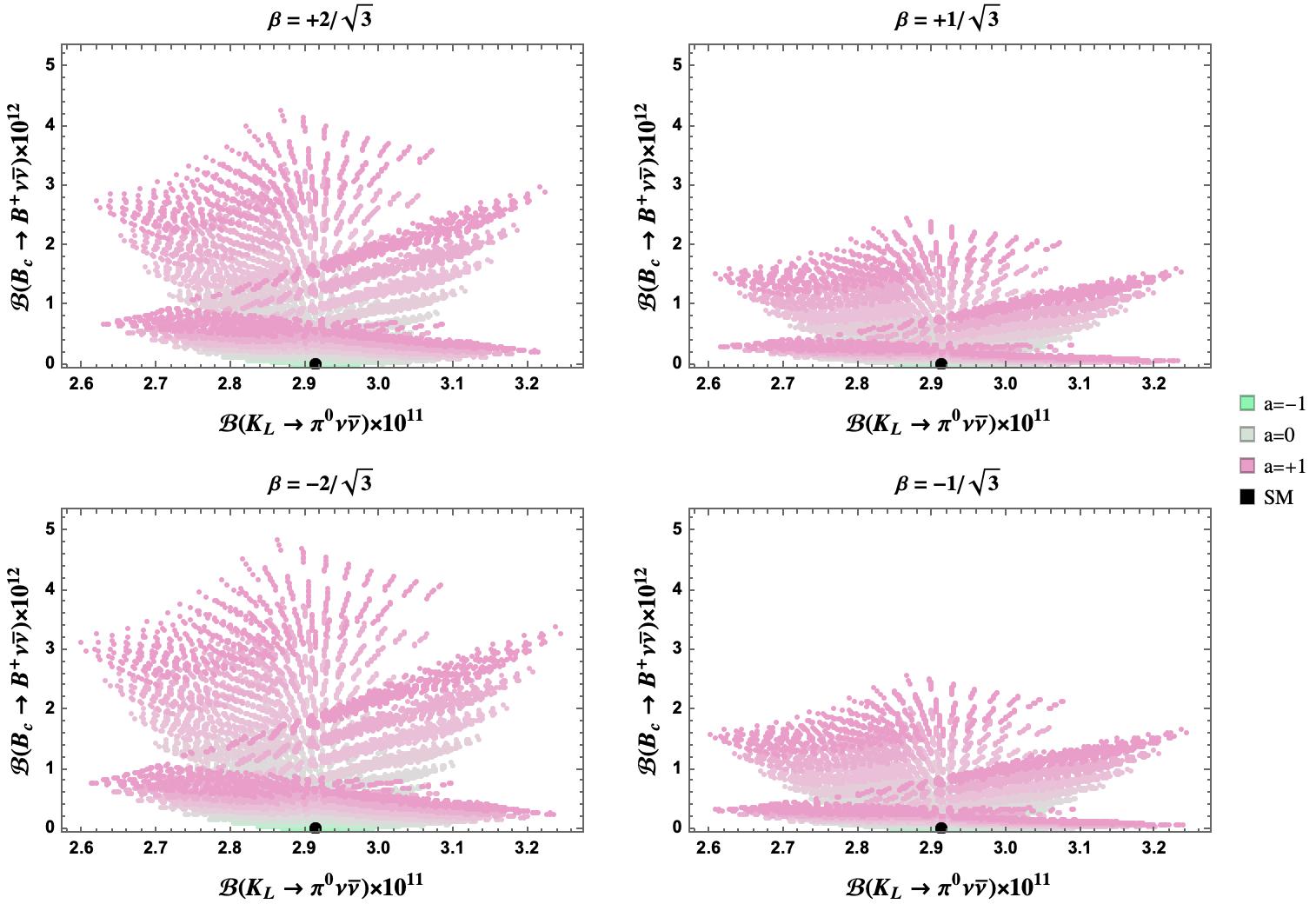}\\
    \caption{\small Correlations between  ${\cal B}(B_c \to B^+ \nu \bar \nu)$ and ${\cal B}(K_L \to \pi^0 \nu \bar \nu)$ in the 331 model with  parameters as in Fig.~\ref{plKpiuBu}. The black dot is the SM result}\label{plK0Bu}
\end{center}
\end{figure}

\begin{figure}[]
\begin{center}
\includegraphics[width = 0.8\textwidth]{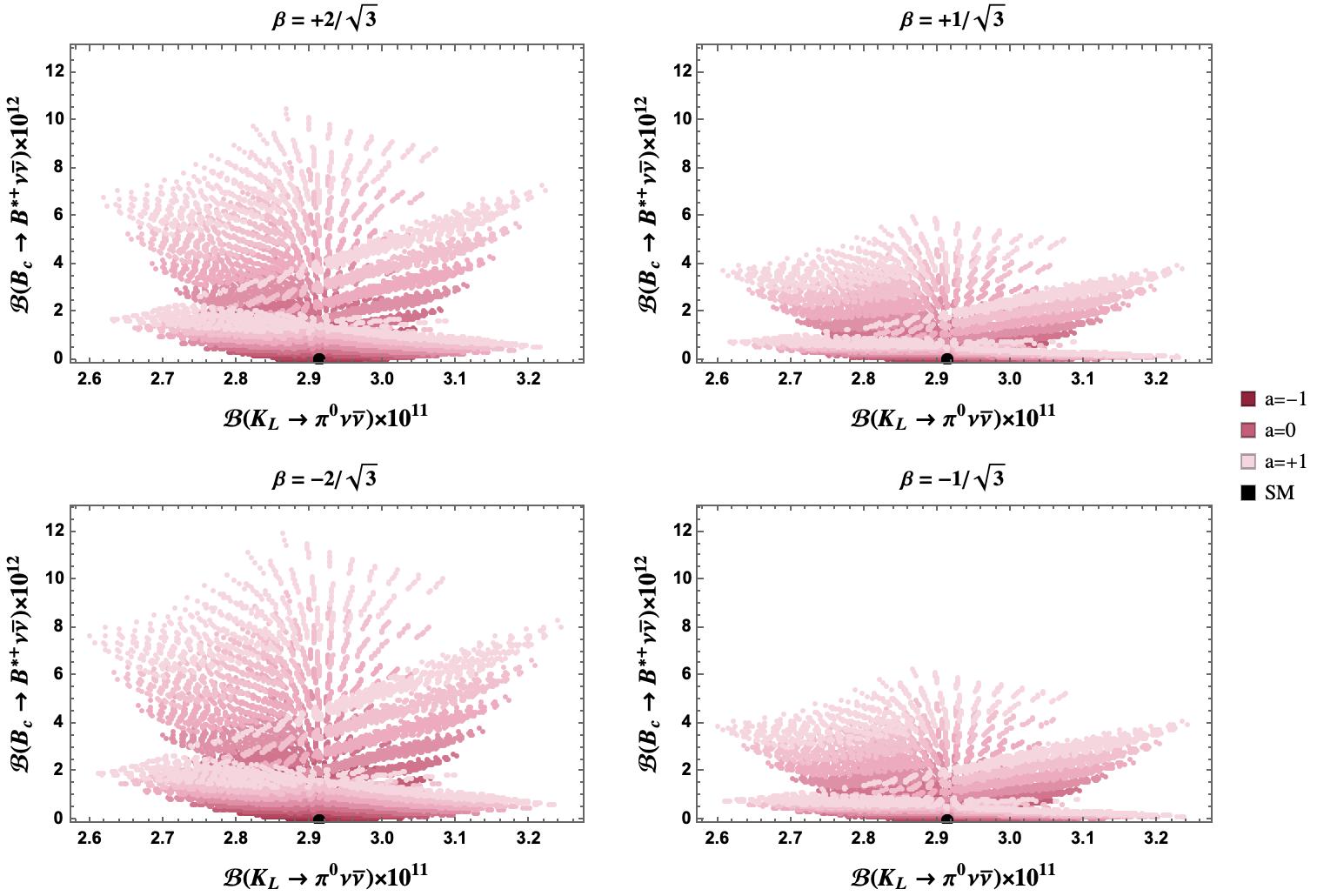}\\
    \caption{\small Correlations between  ${\cal B}(B_c \to B^{*+} \nu \bar \nu)$ and ${\cal B}(K_L \to \pi^0 \nu \bar \nu)$ in the 331 model  with  parameters  as in Fig.~\ref{plKpiuBu}.  The black dot is the SM result.}\label{plK0Bustar}
\end{center}
\end{figure}

In Figs.~\ref{plKpiuBu} and  \ref{plKpiuBustar} we show the correlations between ${\cal B}(B_c \to B^{(*)+} \nu {\bar \nu})$ and  ${\cal B}(K^+ \to \pi^+ \nu {\bar \nu})$,  and in Figs.~\ref{plK0Bu} and  \ref{plK0Bustar}  the correlations with  ${\cal B}(K_L \to \pi^0 \nu {\bar \nu})$. 
The $Z^\prime$ mass is set to $M_{Z^\prime}=1$ TeV, the results for heavier $Z^\prime$ can be obtained by a simple rescaling.  For each $\beta=\pm 1/{\sqrt 3}$,  $\pm 2/{\sqrt 3}$,  and the parameters ${\tilde s}_{13},\,\delta_1$,   ${\tilde s}_{23},\, \delta_2$ are varied in their allowed regions in Fig.~\ref{oases}.
In  each plot the sliding colors represent nine values of the $Z-Z^\prime$ mixing parameter $a$ in the range $[-1,\,1]$ (the colors corresponding to $a=-1,\,0,\,1$ are indicated in the legends).  Only for $a=-1$ are the results  compatible with the SM. For $a=1$ the branching fractions sizably deviate from the SM prediction, and the largest  enhancement of the $B_c$ modes corresponds to a suppression of the kaon  modes with respect to the SM. 

 \subsection{Correlations with  dineutrino $B$ decays}
\begin{figure}[t]
\begin{center}
\includegraphics[width = 0.95\textwidth]{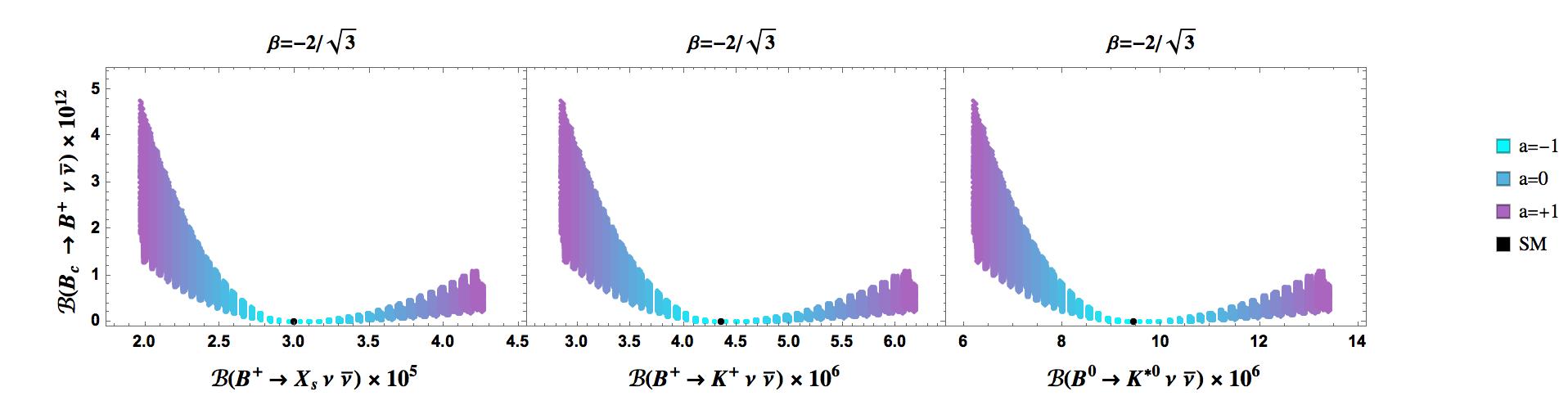}\\
    \caption{\small Correlations between   ${\cal B}(B_c \to B^{+} \nu \bar \nu)$  and   ${\cal B}(B^+ \to  X_s \nu \bar \nu)$  (left), ${\cal B}(B^+ \to  K^+ \nu \bar \nu)$ (middle) and   ${\cal B}(B^0 \to  K^{*0} \nu \bar \nu)$ (right) for  $\beta= -\frac{2}{\sqrt 3}$,  $M_{Z^\prime}=1$ TeV. The black dot indicates the SM result.} \label{plbtos2m}
\end{center}
\end{figure}

\begin{figure}[b]
\begin{center}
\includegraphics[width = 0.95\textwidth]{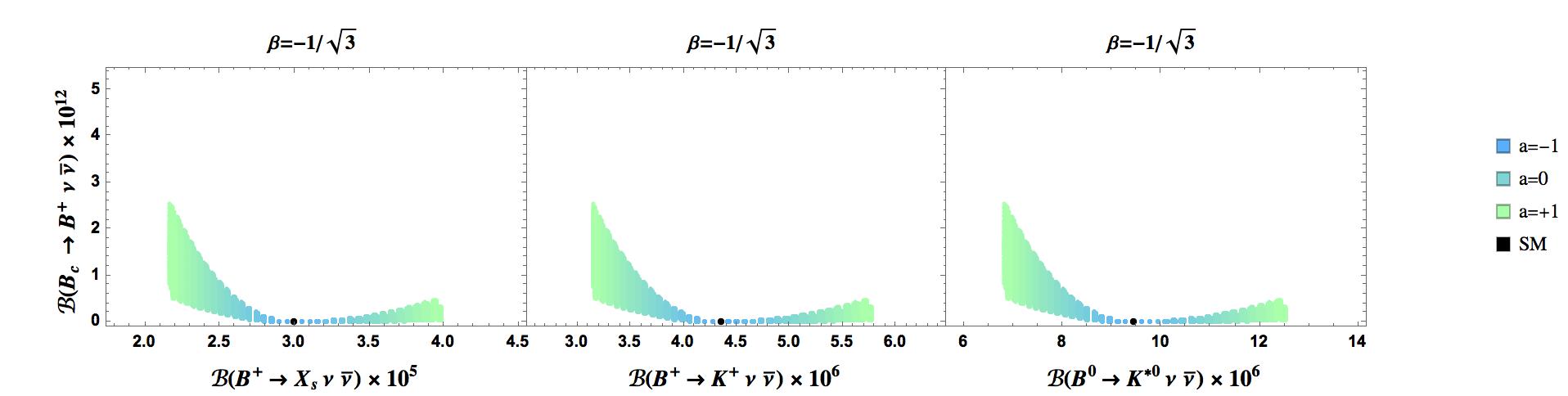}\\
    \caption{\small Correlations between   ${\cal B}(B_c \to B^{+} \nu \bar \nu)$  and   ${\cal B}(B^+ \to  X_s \nu \bar \nu)$  (left), ${\cal B}(B^+ \to  K^+ \nu \bar \nu)$ (middle) and   ${\cal B}(B^0 \to  K^{*0} \nu \bar \nu)$ (right), for  $\beta= -\frac{1}{\sqrt 3}$,  $M_{Z^\prime}=1$ TeV. The black dot indicates the SM result.} \label{plbtos1m}
\end{center}
\end{figure}

Let us consider  the modes $B  \to  M_s \nu {\bar \nu}$   ($M_s=X_s,\,K,\,K^*$). The NP effects  in scenarios with a  $Z^\prime$ with tree-level flavor-changing couplings  only to left-handed fermions can be expressed in the form \cite{Altmannshofer:2009ma,Buras:2012dp}:
 \bea
 \mathcal{B}(B\to M_s \nu \bar \nu) =
 \mathcal{B}(B\to M_s \nu \bar \nu)_{\rm SM} \times \varrho^2~ , \label{eq:BKnn}
 \eea
with
 \be\label{etaepsilon}
 \varrho = \frac{ |X(B_s)|}{ X(x_t)}~ \,\, 
 \ee
and $X(B_s)$  in Eqs.~(\ref{XNP}) and (\ref{XLB}). 
The  SM terms are   \cite{Altmannshofer:2009ma,Buras:2014fpa,Buras:2020xsm}
\bea
\mathcal{B}(B \to X_s \nu \bar \nu)_{\rm SM} &=&(3.0 \pm 0.3) \times 10^{-5} \,\,  \label{BXsSM} \\
 \mathcal{B}(B^+\to K^+ \nu \bar \nu)_{\rm SM} &=&(4.35 \pm 0.59) \times 10^{-6} \label{BKSM} \\
 \mathcal{B}(B^0\to K^{*0} \nu \bar \nu)_{\rm SM} &=&(9.44 \pm 0.89) \times 10^{-6} \,. \label{BKstarSM} 
 \eea
They can be compared to the experimental upper bounds  (at $90\%$ C.L.)  \cite{BaBar:2013npw,Grygier:2017tzo}
\bea
\mathcal{B}(B^+\to K^+ \nu \bar \nu)_{\rm exp} &\le&1.6  \times 10^{-5} \label{BKexp} \\
 \mathcal{B}(B\to K^{*} \nu \bar \nu)_{\rm exp} &\le&2.7 \times 10^{-5} \,\,.\label{BKstarexp} 
 \eea
 The correlations between ${\cal B}(B_c \to B^+ \nu {\bar \nu})$ and ${\cal B}(B \to (X_s,K,K^*)\nu {\bar \nu})$ for $\beta=\displaystyle{- \frac{2}{\sqrt{3}},\,- \frac{1}{\sqrt{3}}}$  are  in Figs.~\ref{plbtos2m} and \ref{plbtos1m}.
The correlations   for  the vector $B_c \to B^{*+} \nu {\bar \nu}$ mode have the same pattern, differing only for the ${\cal B}(B_c \to B^{*+} \nu {\bar \nu})$ scale factor.
The sliding colors describe the variation of $a \in[-1,\,1]$;  the 331 result is compatible with the SM for  $a=-1$.
The largest enhancements of ${\cal B}(B_c \to B^+ \nu {\bar \nu})$ correspond to a suppression of  $b \to s \nu {\bar \nu}$  with respect to the SM. 
 
\section{Conclusions}
 Null tests, like rare FCNC charm decays, are useful to  investigate  the existence of phenomena beyond the Standard Model. The  $B_c \to B^{(*)+} \nu \bar \nu$ modes are  predicted within the SM with branching ratios not exceeding  ${\cal O}(10^{-16})$ and can be used for null tests.
Exploiting the relations between the Wilson coefficients of the $c \to u \nu {\bar \nu}$ and  $c \to u \ell^+ \ell^-$ low-energy Hamiltonian  obtained from the SMEFT, together with the  experimental bounds on the charged dilepton $c \to u$ processes, we have derived  the largest enhancement  for the dineutrino modes in generic NP scenarios, finding  branching fractions  up to ${\cal O}(10^{-6})$. Specific NP scenarios predict   smaller  effects. To investigate this point we have predicted the branching fractions for these processes in the 331 model. The reason to consider this framework is that in this model the NP parameters entering in FCNC  charm decays are the same that govern the $B,\,B_s,\,K$ FCNC transitions. This provides nontrivial correlations among the observables in the various systems. 
 In the 331 models  the effective $c \to u \nu {\bar \nu}$ Hamiltonian comprises only one operator, as in the SM, with a modification of the  Wilson coefficient  enhancing  ${\cal B}(B_c \to B^{(*)+} \nu \bar \nu)$  up to ${\cal O}(10^{-11})$.
Moreover, in this model a correlation  with down-type quark dineutrino processes can be established. We have found  that the largest branching fractions  correspond to  $\beta=-2/\sqrt{3}$  and are  anticorrelated with  $K^+ \to \pi^+ \nu {\bar \nu}$, $K_L \to \pi^0 \nu {\bar \nu}$ and $B \to \{X_s,\,K,\,K^*\} \nu {\bar \nu}$.

\section*{Acknowledgements}
We thank A.J. Buras for enlightening discussions.
This study has been  carried out within the INFN project (Iniziativa Specifica) QFT-HEP.
\newpage
\appendix
\numberwithin{equation}{section}
\section{LONG-DISTANCE CONTRIBUTIONS TO $B_c \to B^{(*)+} \nu \bar \nu$}\label{sec:ld} 
We estimate the main long-distance contributions to $B_c \to B^{(*)+} \nu \bar \nu$ represented by the processes 
$B_c \to B^{(*)+} V^0 \to  B^{(*)+} \nu \bar \nu$, with $V^0=\rho^0, \omega, \phi$ \cite{Burdman:2001tf}. The nonleptonic color suppressed  
$B_c \to B^{(*)+} V^0$  amplitude can be estimated using naive factorization in terms of the $B_c \to B^{(*)}$ form factors. The $V^0 \to \nu_\ell \bar \nu_\ell$ amplitude  involves the hadronic matrix elements
\be
\langle 0| \bar q \gamma^\mu (g_V^q - g_A^q \gamma_5) q | V^0(q, \epsilon)\rangle \label{eq:frho}
\ee
with $g_{V,A}^q$ the vector and axial-vector  couplings constants of the neutral current for  quarks. Actually,  \eqref{eq:frho}
 takes contribution only from the vector quark current. These matrix elements can be obtained from the $V^0$ matrix element
of the em current $\dd J^\mu_{em}=\sum_{q} e_q \bar q \gamma^\mu q$, with $e_q$ the quark charges, parametrized as
\be
\langle 0| J^\mu_{em}| V^0(q, \epsilon)\rangle =\frac{m^2_{V^0}}{f_{V^0}} \epsilon^\mu \, \, .
\ee
Using  the $V^0$ masses, widths and  $V^0 \to e^+ e^-  (\mu^+ \mu^-)$ branching fractions \cite{Zyla:2020zbs}, we have 
 $f_{\rho^0}=4.99\pm0.03$ $(5.08\pm0.16)$,
  $f_{\omega}=16.50\pm0.25$  $(16.49\pm2.01)$,
   $f_{\phi}=13.51\pm0.22$    $(13.78\pm0.51)$. The results for the LD contribution   
${\cal B}(B_c \to B^+ \nu \bar \nu)|_{LD}\simeq 1.0 \times 10^{-16}$,  ${\cal B}(B_c \to B^{*+} \nu \bar \nu)|_{LD}\simeq  9.8 \times 10^{-17}$
 confirm the role of  $B_c \to B^{(*)} \nu \bar \nu$ as null tests of  the SM.

\newpage
\bibliographystyle{apsrev4-1}
\bibliography{refFFP5}
\end{document}